\documentclass[12pt]{article}

\usepackage{graphics}
\usepackage{epsfig}
\usepackage{cite}
\input epsf

\title{Investigation of beauty production and parton shower effects at LHC}

\author{H.~Jung$^{1,2}$, M.~Kraemer$^1$, A.V.~Lipatov$^3$, N.P.~Zotov$^3$}

\begin{document}
\maketitle

\vspace*{-7.5cm}
\begin{flushright}
DESY 11-180\\
November 2011 \\
\end{flushright}
\vspace*{+4.5cm}

\begin{center}

{\it $^1$DESY, Hamburg, Germany\\[3mm]
$^2$ CERN, Geneva, Switzerland\\
University of Antwerp, Antwerp, Belgium\\[3mm]
$^3$D.V.~Skobeltsyn Institute of Nuclear Physics,\\ 
M.V. Lomonosov Moscow State University, Russia}\\[3mm]

\end{center}

\vspace{0.5cm}

\begin{center}

{\bf Abstract }

\end{center}

We present hadron-level predictions from the Monte Carlo generator {\sc Cascade} and 
parton level calculations of open $b$ quark, $b$-flavored hadron 
and inclusive $b$-jet production in the framework of 
the $k_T$-factorization QCD approach for the LHC energies. 
The unintegrated gluon densities in a proton are determined using the CCFM 
evolution equation and the Kimber-Martin-Ryskin (KMR) prescription.
Our predictions are compared with the first data taken by 
the CMS and LHCb collaborations at $\sqrt s = 7$~TeV.
We study the theoretical uncertainties of our calculations and investigate the 
effects coming from parton showers in initial and final states.
The special role of initial gluon transverse momenta in description
of the data is pointed out.

\vspace{0.8cm}

\noindent
PACS number(s): 12.38.-t, 13.85.-t

\vspace{0.5cm}

\section{Introduction} \indent 

Beauty production at high energies is subject of intense 
studies from both theoretical and experimental 
points of view since events containing $b$ quarks present an important 
background to many of the searches at the LHC.
From the theoretical point, the dominant production mechanism is believed 
to be quark pair production through the gluon-gluon fusion subprocess and
therefore these processes provide an opportunity to test the different 
predictions based on Quantum Chromodynamics (QCD).
The present note is motivated by the recent measurements\cite{1,2,3,4,5,6}
of beauty production performed by the 
CMS and LHCb collaborations at the LHC energy $\sqrt s = 7$~TeV. 
It was observed\cite{2,3,4} that the data on $B^+$, $B^0$ meson and open $b$ quark production
tend to be higher than the \textsc{MC@NLO}\cite{7,8} predictions. There are no
predictions which describe reasonably well the angular correlations between $b$-flavored hadrons
measured\cite{5} by the CMS collaboration.
On the other hand, the measurements of transverse momenta and rapidity
distributions of beauty hadrons\cite{1} and inclusive
$b$-jets\cite{6} are reasonably well described 
by the \textsc{MC@NLO}. 

In the framework of the $k_T$-factorization approach of QCD\cite{9}, 
  heavy quark 
production has been studied (for previous results see\cite{10,11,12,13,14,15,16}).
In our previous paper\cite{16} we have obtained a good agreement between the Tevatron data on 
the open $b$ quarks, $b \bar b$ di-jets, 
$B^+$ and several $D$ mesons (or rather muons from their semileptonic decays) 
production with the predictions coming from $k_T$-factorization and we have investigated 
the role of initial and final state parton showers.
We have shown that a good description of the specific angular correlations between the final-state 
particles is obtained in Monte Carlo event generator \textsc{Cascade}\cite{17} 
once the higher order process $gg^* \to gg$ with subsequent $g\to b\bar{b}$ splitting is included, which is not discussed  here.

Based on these results, here we give 
a systematic analysis of the recent CMS and LHCb data\cite{1,2,3,4,5,6} 
on beauty production in the framework of $k_T$-factorization\footnote{See also\cite{18}}. 
Following to\cite{16}, we produce the calculations in two ways:
we perform numerical parton-level calculations (labeled as LZ) as well as 
calculations with the full hadron level Monte Carlo event generator 
\textsc{Cascade} and compare both with the measured cross 
sections\footnote{In addition to the comparison of CASCADE predictions 
with the data in\cite{4} we present here further studies.}.
We investigate the influence of parton showers in 
initial and final states for the description of LHC data.
Specially we concentrate on the angular correlations between the produced
$b$-flavored hadrons measured by the CMS collaboration\cite{5}
which are important for our understanding of production dynamics\cite{14,15,16}.
Finally, we study the different sources of theoretical uncertainties, i.e. uncertainties 
connected with the gluon evolution scheme, 
heavy quark mass, hard scale 
of partonic subprocess and the heavy quark fragmentation functions. 

The outline of our paper is the following. In Section~2 we 
recall the basic formulas of the $k_T$-factorization approach with a brief 
review of calculation steps. In Section~3 we present the numerical results
of our calculations and a discussion. Section~4 contains our conclusions.

\section{Theoretical framework} \indent 

In the present note we follow the approach
described in the earlier publications\cite{14,15,16}.
For the reader's convenience, we only briefly recall
here main points of the theoretical scheme.
The cross section of heavy quark production in $pp$ collisions
at high energies in the $k_T$-factorization approach
is calculated as a convolution of the off-shell (i.e. $k_T$-dependent)
partonic cross section and the unintegrated gluon 
distributions in a proton. It can be presented in the following form:
$$
  \displaystyle \sigma (p p \to Q\bar Q \, X) = \int {1\over 16\pi (x_1 x_2 s)^2 } {\cal A}(x_1,{\mathbf k}_{1T}^2,\mu^2) {\cal A}(x_2,{\mathbf k}_{2T}^2,\mu^2) |\bar {\cal M}(g^* g^* \to Q\bar Q)|^2 \times \atop
  \displaystyle  \times d{\mathbf p}_{1T}^2 d{\mathbf k}_{1T}^2 d{\mathbf k}_{2T}^2 dy_1 dy_2 {d\phi_1 \over 2\pi} {d\phi_2 \over 2\pi}, \eqno (1)
$$

\noindent 
where ${\cal A}(x,{\mathbf k}_{T}^2,\mu^2)$ is the
unintegrated gluon distribution in a proton, 
$|\bar {\cal M}(g^* g^* \to Q\bar Q)|^2$ is the 
off-shell (i.e. depending on the initial gluon virtualities 
${\mathbf k}_{1T}^2$ and ${\mathbf k}_{2T}^2$) matrix element squared 
and averaged over initial gluon 
polarizations and colors, and 
$s$ is the total center-of-mass energy.
The produced heavy quark $Q$ and anti-quark $\bar Q$ have the 
transverse momenta ${\mathbf p}_{1T}$ and
${\mathbf p}_{2T}$ and the center-of-mass rapidities $y_1$ and $y_2$.
The initial off-shell gluons have a fraction $x_1$ and $x_2$ 
of the parent protons longitudinal 
momenta, non-zero transverse momenta ${\mathbf k}_{1T}$ and 
${\mathbf k}_{2T}$ (${\mathbf k}_{1T}^2 = - k_{1T}^2 \neq 0$, 
${\mathbf k}_{2T}^2 = - k_{2T}^2 \neq 0$) and azimuthal angles
 $\phi_1$ and $\phi_2$. 
The analytic expression for the 
$|\bar {\cal M}(g^* g^* \to Q\bar Q)|^2$ can be found, for example, in\cite{9,13}.

The unintegrated gluon distributions in a 
proton ${\cal A}(x,{\mathbf k}_{T}^2,\mu^2)$ involved in~(1) can be obtained from the analytical or 
numerical solutions of the Balitsky-Fadin-Kuraev-Lipatov (BFKL)\cite{19} or 
Ciafaloni-Catani-Fiorani-Marchesini (CCFM)\cite{20} evolution equations. 
As in\cite{12}, we have tested 
a few different sets. First of them, CCFM set A0 has been obtained in\cite{21} from the CCFM equation
where all input parameters have been fitted to describe the proton structure function $F_2(x, Q^2)$.
Equally good fit was obtained using different values for the soft cut 
and a different value for the width of the intrinsic ${\mathbf k}_{T}$ distribution 
(CCFM set B0). 
Also we will use the unintegrated gluon densities in a proton
taken in the Kimber-Martin-Ryskin form\cite{22}. The KMR approach is a formalism to construct the 
unintegrated parton distributions from well-known conventional ones. 
For the input, we have used the standard MSTW'2008~(LO)\cite{23}
(in LZ calculations) and MRST 99\cite{24} (in \textsc{Cascade}) sets.

\section{Numerical results} \indent

The unintegrated gluon distributions to be used in the cross section (1) depend on
the renormalization and factorization scales $\mu_R$ and $\mu_F$.
Following to\cite{16}, in the numerical calculations we set 
$\mu_R^2 = m_Q^2 + ({\mathbf p}_{1T}^2 + {\mathbf p}_{2T}^2)/2$,
$\mu_F^2 = \hat s + {\mathbf Q}_T^2$, where ${\mathbf Q}_T$ is the 
transverse momentum of the initial off-shell gluon pair,
$m_c = 1.4 \pm 0.1$~GeV, $m_b = 4.75 \pm 0.25$~GeV.  We use the LO formula 
for the coupling $\alpha_s(\mu_R^2)$ with $n_f = 4$ active quark flavors
at $\Lambda_{\rm QCD} = 200$~MeV, such 
that $\alpha_s(M_Z^2) = 0.1232$. 

We are in position to present our numerical results.
The CMS collaboration has measured $B^+$ and $B^0$ meson cross sections
in the kinematic range $p_T(B^+) > 5$~GeV, $|y(B^+)| < 2.4$\cite{2} and
$p_T(B^0) > 5$~GeV, $|y(B^0)| < 2.2$\cite{3}, respectively.
The measurements of decay muon cross sections have been performed\cite{4} for
$p_T(\mu) > 6$~GeV and $|\eta(\mu)| < 2.1$.
The LHCb collaboration have measured\cite{1} the pseudorapidity distribution of 
$b$-hadrons in forward region $2 < \eta(H_b) < 6$,
where $H_b$ is any $b$-flavored hadron.
In our calculations the fragmentation of $b$ quarks into a $b$ hadrons is 
described with the Peterson fragmentation
function\cite{25} with default value of shape parameter $\epsilon_b = 0.006$.
To produce muons from $b$ quarks in the LZ calculations, we first convert $b$ quarks into $b$ hadrons
and then simulate their semileptonic decay according to the standard electroweak theory.
The branching fractions of $b \to B^+$, $b \to B^0$, $b \to \mu $ as well as the cascade 
decay $b\to c\to \mu$ are taken from\cite{26}.
The CMS collaboration has presented preliminary data\cite{6} on the 
inclusive $b$-jet production at the $\sqrt s = 7$~TeV.
The cross sections have been determined in four $b$-jet rapidity regions,
namely $|y| < 0.5$, $0.5 < |y| < 1$, $1 < |y| < 1.5$ and $1.5 < |y| < 2$. 
The $b$-jets in the \textsc{Cascade} calculations are reconstructed with the 
anti-$k_t$ cone algorithm\cite{27} (using the {\sc Fastjet} package\cite{28,29}) 
with radius $R = 0.5$.

The results of our calculations are shown in 
Figs.~1 --- 6 in comparison with the data.
We obtain a good description of the data when using 
the CCFM-evolved (namely, A0) gluon distribution in LZ calculations.
The shape and absolute normalization of measured $b$-flavored hadron 
cross sections at forward rapidities
are reproduced well (see Fig.~4). 
The KMR and CCFM B0 predictions are somewhat below the data.
In contrast with $b$ hadron and decay muon cross sections, 
the results for inclusive $b$-jet production 
based on the CCFM and KMR gluons are very similar to each other and
a reasonable description of the data is obtained by all unintegrated 
gluon distributions under consideration.

The {\sc Cascade} predictions 
tend to lie slightly below the LZ ones and are rather
close to the {\textsc MC@NLO} calculations\cite{7,8} (not shown). 
The observed difference between the LZ and {\sc Cascade} is not surprising
and connected with the missing parton shower effects in the LZ evaluations.
The influence of such effects is clearly demonstrated in Fig.~7, where we 
show separately the results of \textsc{Cascade} calculations  
without parton shower, with only initial state, with only final state and with 
both initial and final state parton showers. 
One can see that without
initial and final state parton showers, the \textsc{Cascade} predictions
are very close to the LZ ones.
The similar situation was pointed out previously\cite{16} at for Tevatron energies.
We have checked that the LZ and \textsc{Cascade} predictions coincide at parton level.

\begin{table}
\begin{center}
\begin{tabular}{|l|c|c|c|}
\hline
   & & &\\
  Source & $\sigma(B^+)$ & $\sigma(B^0)$ & $\sigma(\mu)$\\
   & & &\\
\hline
   & & &\\
   CMS data [$\mu$b] & $28.1 \pm 2.4 \pm 2.0 \pm 3.1$ & $33.2 \pm 2.5 \pm 3.5$ & $1.32 \pm 0.01 \pm 0.30 \pm 0.15$\\
   & & &\\
   \hline
   & & &\\
   A0 (LZ/\textsc{Cascade}) & 32.7/24.5 & 31.4/24.3 & 1.31/0.96 \\
   & & &\\
   B0 (LZ/\textsc{Cascade}) & 23.6/18.8 & 22.5/20.4 & 0.98/0.72 \\
   & & &\\
   KMR (LZ/\textsc{Cascade}) & 16.7/13.1 & 15.8/12.4 & 0.91/0.59 \\
   & & &\\
   \textsc{MC@NLO} \protect\cite{2,3,4} & 19.1 & 25.2 & 0.95 \\
   & & &\\
   \textsc{Pythia}  \protect\cite{2,3,4} & 36.2 & 49.1 & 1.9 \\
   & & &\\
   \hline
\end{tabular}
\end{center}
\caption{The $b$-flavored hadron production cross section in 
  $pp$ collisions at $\sqrt s = 7$~TeV.}
\label{table_pdfs}
\end{table}

In order to study the dependence of our predictions on the quark-to-hadron fragmentation function,
we repeated our calculations with the shifted value of the Peterson 
shape parameter $\epsilon_b = 0.003$, which is 
is often used in the NLO pQCD calculations.
Additionally, we have applied the non-perturbative fragmentation functions 
proposed in\cite{30,31,32} where the input parameters 
were determined in\cite{31,32} by a fit to LEP2 data. The results of our calculations
are shown in Fig.~8. We find that the predicted cross sections 
in the considered kinematic region are larger for smaller values of the 
parameter $\epsilon_b$ or if the fragmentation function from\cite{30,31,32} is used.
Thus, the LHC data lie within the band of theoretical uncertainties. 

The visible cross sections of $b$-flavored hadrons and $b$-decay muons
are listed in Table~1 in comparison with the CMS data~\cite{2,3,4}. In Table~2
the systematic uncertainties of our calculations are summarized.
To estimate the uncertainty coming from the renormalization scale $\mu_R$, we used 
the CCFM set A0$+$ and A0$-$ instead of the default density function A0.
These two sets represent a variation of the scale used in $\alpha_s$ in the 
off-shell matrix element. The A0$+$ stands for a variation of $2\,\mu_R$, 
while set A0$-$ reflects $\mu_R/2$. 
We observe a deviation of roughly $13\%$ for set A0$+$.
The uncertainty coming from set A0$-$ is generally smaller and negative.
The dependence on the $b$-quark mass is investigated by variation of $b$-quark mass of  $m_b=4.75$~GeV by $\pm0.25$ GeV. 
The calculated $b$-quark cross sections vary by $\sim \pm 6 \%$.

\begin{table}
\begin{center}
\begin{tabular}{|l|c|}
\hline
   & \\
  Source & $\sigma(pp \to b + X \to \mu + X^\prime,$ $p_T^\mu > 6$~GeV, $|\eta^\mu| < 2.1)$ \\
   & \\
\hline
   & \\
   CCFM set A0 & 0.96 $\mu $b\\
   & \\
   CCFM set A0$+$ & +13\% \\
   & \\
   CCFM set A0$-$ & -2\% \\
   & \\
   $m_b=5.0$ GeV & -7\% \\
   & \\
   $m_b=4.5$ GeV & +6\% \\
   & \\
   $\epsilon_b=0.003$ & +9\% \\
   & \\
   \hline
   & \\
   Total & $\pm^{17\%}_{7\%}$ \\
   & \\
\hline
\end{tabular}
\end{center}
\caption{Systematic uncertainties for beauty total cross section in 
  $p p$ collisions at $\sqrt s = 7$~TeV obtained with \textsc{Cascade}.}
\label{table_uncertainties}
\end{table}

Now we turn to the investigation of
the angular correlations between the produced $b$ hadrons. 
As it was pointed out in\cite{14,15,16}, such observables are very sensitive to the details
of the non-collinear gluon evolution.
The CMS collaboration\cite{5} has measured the $b$-flavored hadron pair production as a function of the angular
separation $\Delta \phi$ between the two reconstructed $b$ hadrons and 
variable $\Delta R = \sqrt{(\Delta \eta)^2 + (\Delta \phi)^2}$ 
for three different event scales, characterised by the leading jet transverse momentum $p_T$,
namely $p_T > 56$~GeV, $p_T > 84$~GeV and $p_T > 120$~GeV.
The kinematic range for the measurements is defined by the requirements 
$p_T(H_b) > 15$~GeV and $|\eta(H_b)| < 2$ for both of $b$-flavored hadrons. 
The leading jet is required to be within $|\eta| < 3.0$.
Our predictions for $\Delta \phi$ and $\Delta R$ distributions are shown in Figs.~9 and 10.
One can see that none of the calculations
fully describes the LHC data and therefore there is a room for further studies.
Note that the predicted shapes of $\Delta \phi$ and $\Delta R$ distributions are very 
different for different unintegrated gluon densities used, as it was expected.
This is in a contrast to the cross sections as a function of transverse
momenta or rapidities where all gluon distributions gave a similar behaviour.
Note also that the measured cross sections at small $\Delta \phi$ or $\Delta R$ are significant. 
Moreover, they exceed the ones observed at large angular separation where 
the two $b$-flavored hadrons are emitted in opposite directions.
This behavior is reproduced by the KMR gluon density only due to different tail at 
large ${\mathbf k}_{T}$ compared to the CCFM-evolved gluon distributions\footnote{A detailed
comparison of KMR and CCFM gluon distributions can be found in\cite{33}.}.
The role of non-zero gluon transverse momentum ${\mathbf k}_{T}$ is clearly illustrated also 
in Fig.~11.
Here the solid histograms correspond to the results obtained according to the 
master formula~(1) and the dotted histograms are obtained by using the same formula and 
without virtualities of the incoming gluons in partonic amplitude. In the last case an 
additional requirement ${\mathbf k}_{1,2 \,T}^2 < \mu_R^2$ is applied.
One can see that the gluon transverse momentum ${\mathbf k}_{T}$
(both in the hard matrix element and in the gluon distribution functions)
is important for description of the LHC data at low $\Delta \phi$ or $\Delta R$. 

\section{Conclusions} \indent 

In this note we analyzed the first data on the beauty production in 
$pp$ collisions at the LHC taken by the CMS and LHCb collaborations.
Our consideration is based on the $k_T$-factorization 
approach supplemented with the CCFM-evolved unintegrated gluon
densities in a proton. The analysis covers the total and differential cross sections 
of $b$-flavored hadrons and muons originating from the semileptonic decays of  beauty quarks as well as
the double differential cross sections of inclusive $b$-jet production. 
Using the full hadron-level Monte Carlo generator {\sc Cascade}, we investigated the 
effects coming from the parton showers in initial and final states.
Different sources of theoretical uncertainties have been studied.

Our LZ predictions with the default set of parameters agree with the data
on the transverse momentum and pseudorapidity distributions of $b$-flavored hadrons
or $b$ quark decay muons. The \textsc{Cascade} predictions tend to slightly 
underestimate the data at central rapidities but the data points still lie within the band of 
theoretical uncertainties. In this case the overall description of the data at a 
similar level of agreement as in the framework of NLO collinear QCD factorization.
Special attention was put on the analysis of specific angular correlations between the 
produced $b$-flavored hadrons measured by the CMS collaboration. 
The description of of $\Delta \phi$ and $\Delta R$ distributions in the
framework of the $k_T$-factorization require further studies.


\section{Acknowledgments} \indent 

We are grateful to comments from V.~Chiochia, G.~Dissertori, W.~Erdmann, A.~Rizzi, V.~Zhukov and
S.~Baranov. 
The authors are very grateful to 
DESY Directorate for the support in the 
framework of Moscow --- DESY project on Monte-Carlo
implementation for HERA --- LHC.
A.V.L. was supported in part by the grant of president of 
Russian Federation (MK-3977.2011.2).
Also this research was supported by the 
FASI of Russian Federation (grant NS-4142.2010.2),
FASI state contract 02.740.11.0244, 
RFBR grant 11-02-01454-a and the RMES (grant the Scientific Research on High Energy Physics).

\newpage

\begin{figure}
\centering
\begin{picture}(16.5,15.)(0.,0.)
\put(0.1,6.59){\epsfig{figure=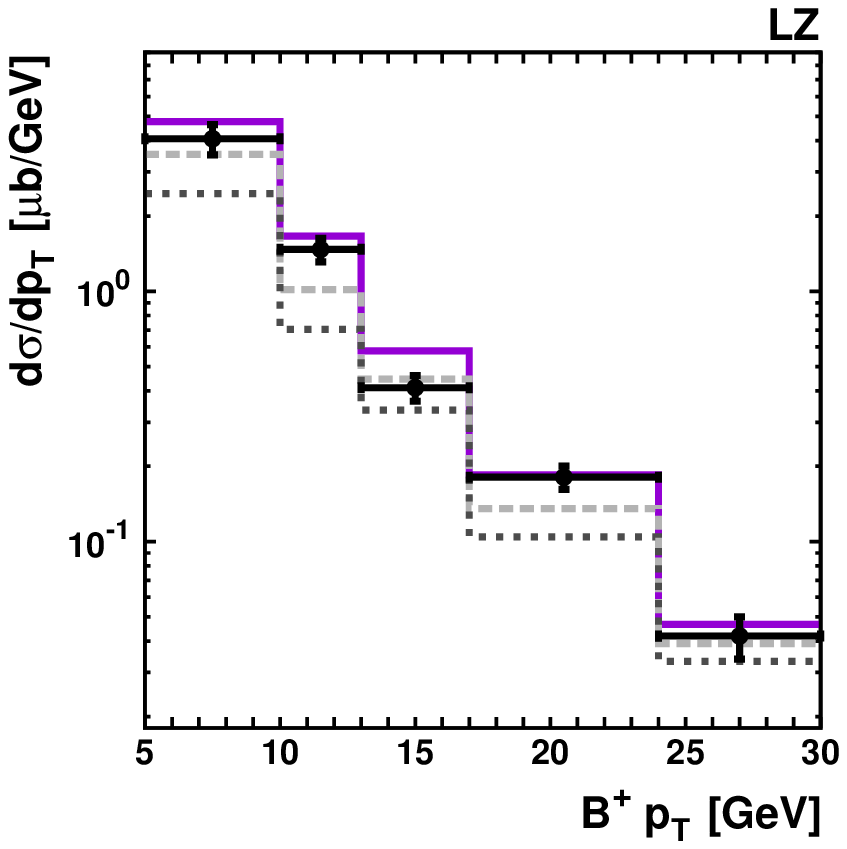, width = 8.1cm}}
\put(9.2,6.){\epsfig{figure=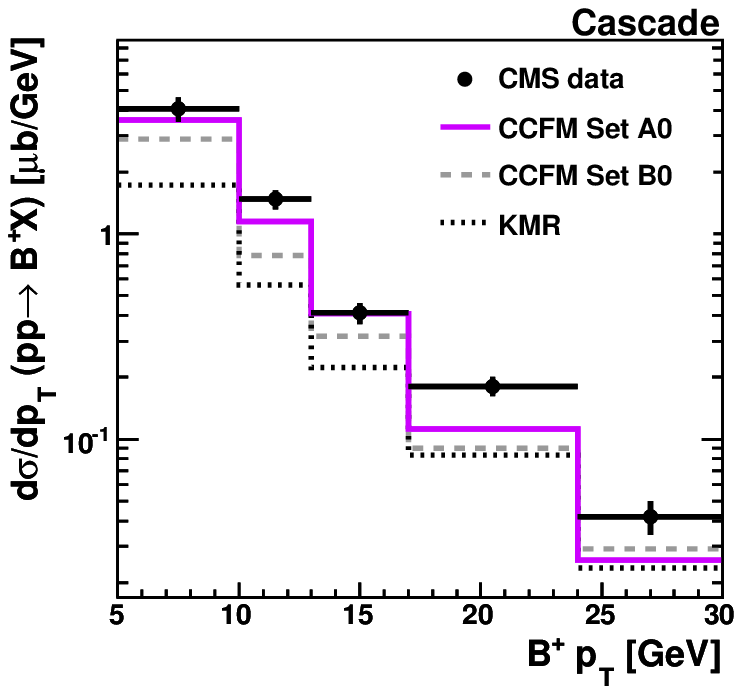, height = 6.3cm, width = 7.3cm}}
\put(0.1,0.59){\epsfig{figure=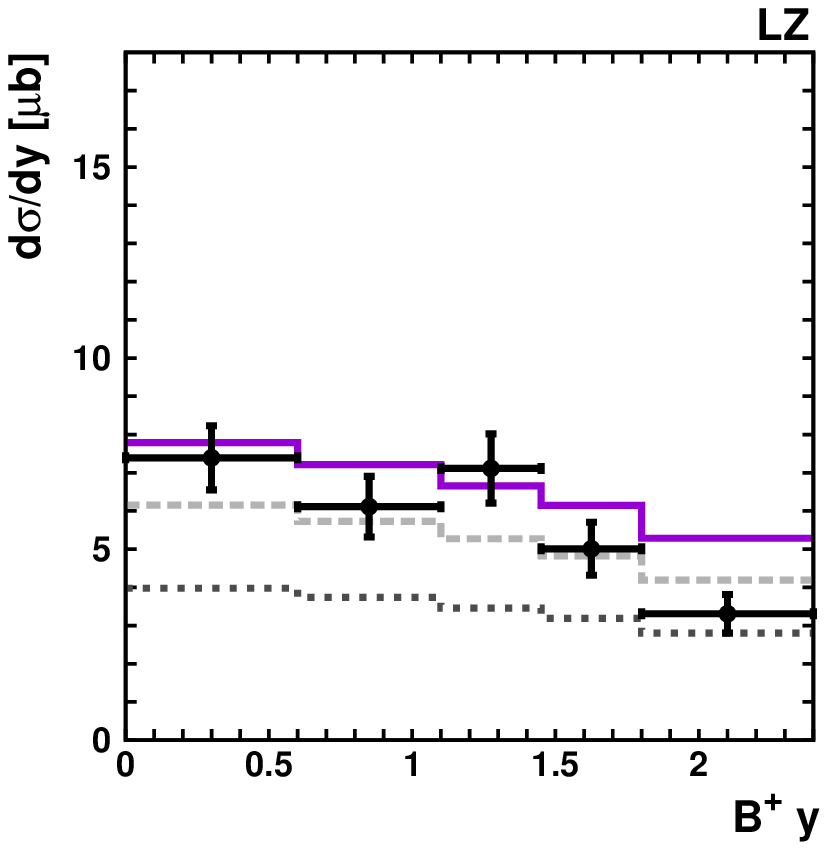, width = 8.1cm}}
\put(9.2,0.0){\epsfig{figure=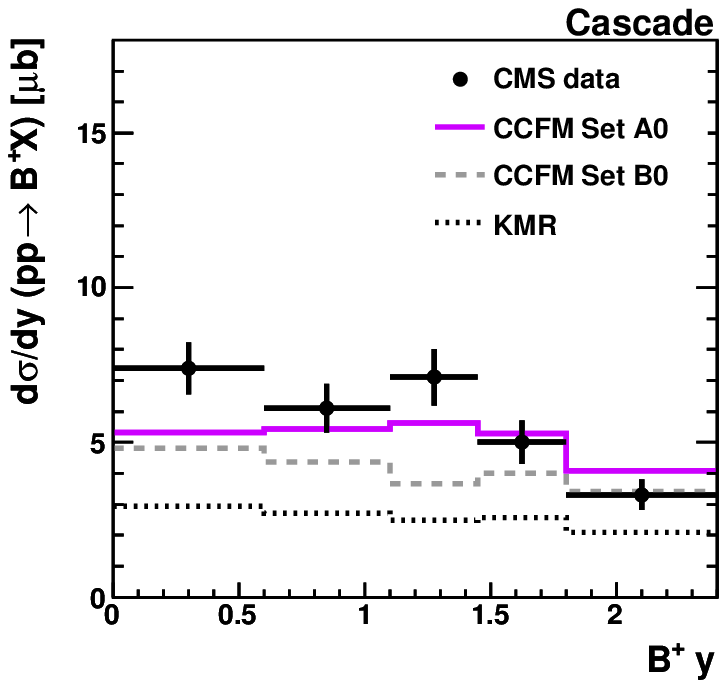, height = 6.3cm, width = 7.3cm}}
\end{picture}
\caption{The transverse momentum and rapidity distributions of $B^+$ meson production at the LHC. 
The kinematical cuts applied are described in the text. 
The solid, dashed and dotted histograms 
correspond to the results obtained with the CCFM set A0, B0
and KMR unintegrated gluon densities.
The first column shows the LZ results while the second one 
depicts the \textsc{Cascade} predictions. The experimental data are from CMS\protect\cite{2}.}
\label{fig1}
\end{figure}

\begin{figure}
\centering
\begin{picture}(16.5,15.)(0.,0.)
\put(0.1,6.59){\epsfig{figure=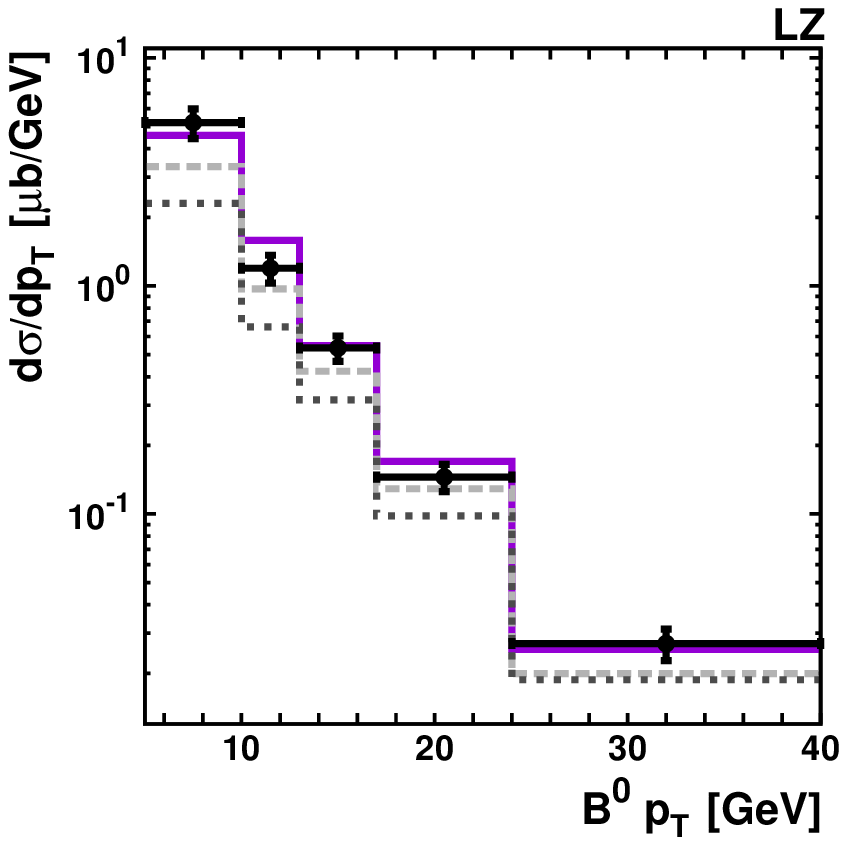, width = 8.1cm}}
\put(9.2,6.){\epsfig{figure=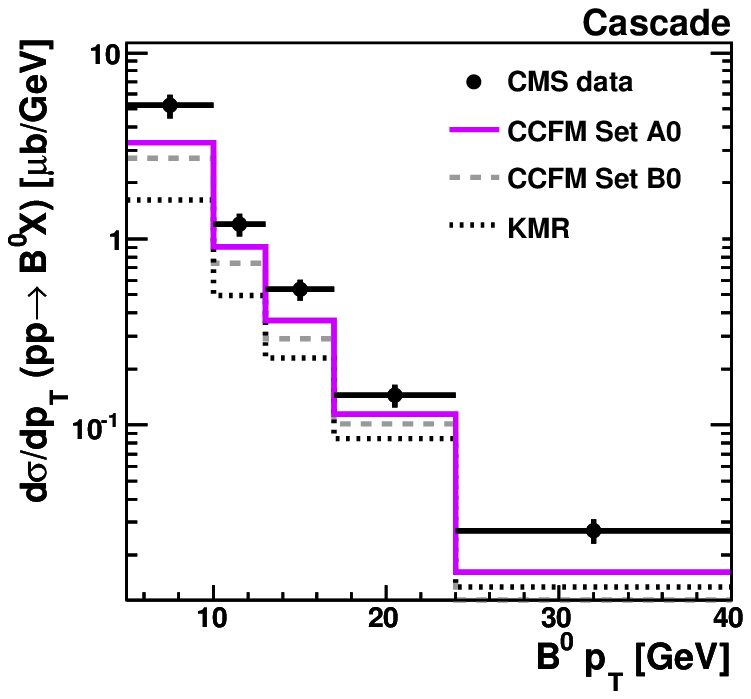, height = 6.3cm, width = 7.3cm}}
\put(0.1,0.59){\epsfig{figure=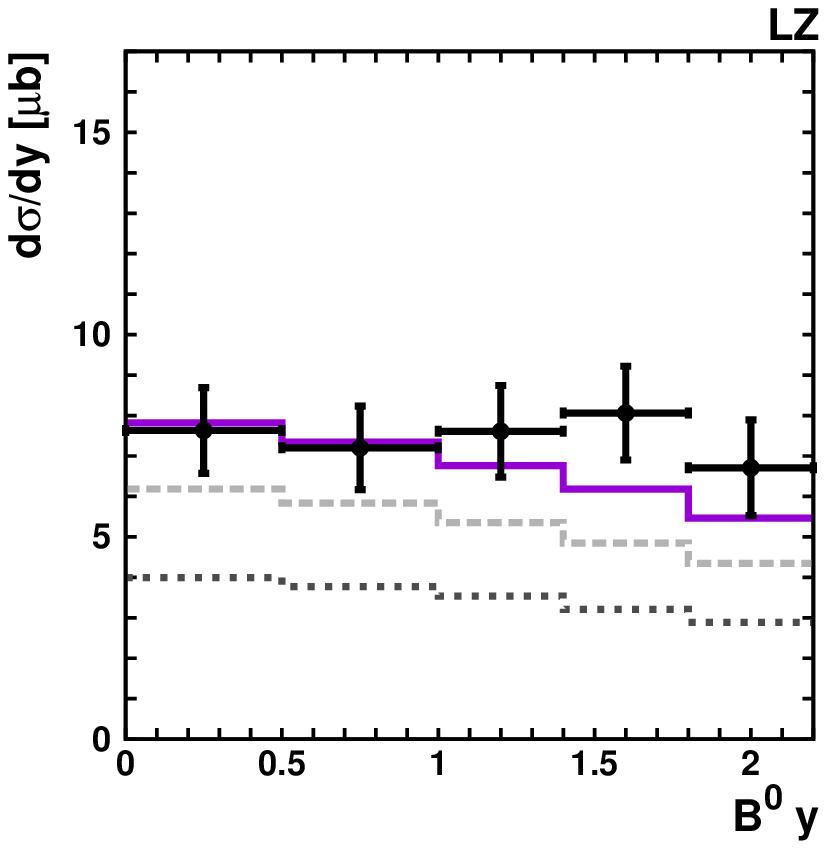, width = 8.1cm}}
\put(9.2,0.0){\epsfig{figure=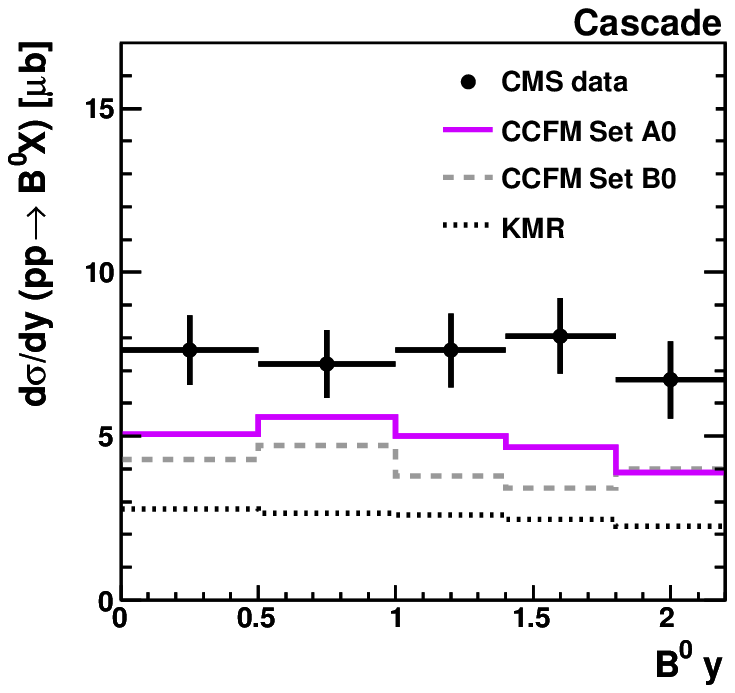, height = 6.3cm, width = 7.3cm}}
\end{picture}
\caption{The transverse momentum and rapidity distributions of $B^0$ meson production at the LHC. 
The kinematical cuts applied are described in the text. The left histograms show 
the LZ numerical results while the right plots depict the \textsc{Cascade} predictions.
Notation of all histograms is the same as in Fig.~1. 
The experimental data are from CMS\protect\cite{3}.}
\label{fig2}
\end{figure}

\begin{figure}
\centering
\begin{picture}(16.5,15.)(0.,0.)
\put(0.1,6.59){\epsfig{figure=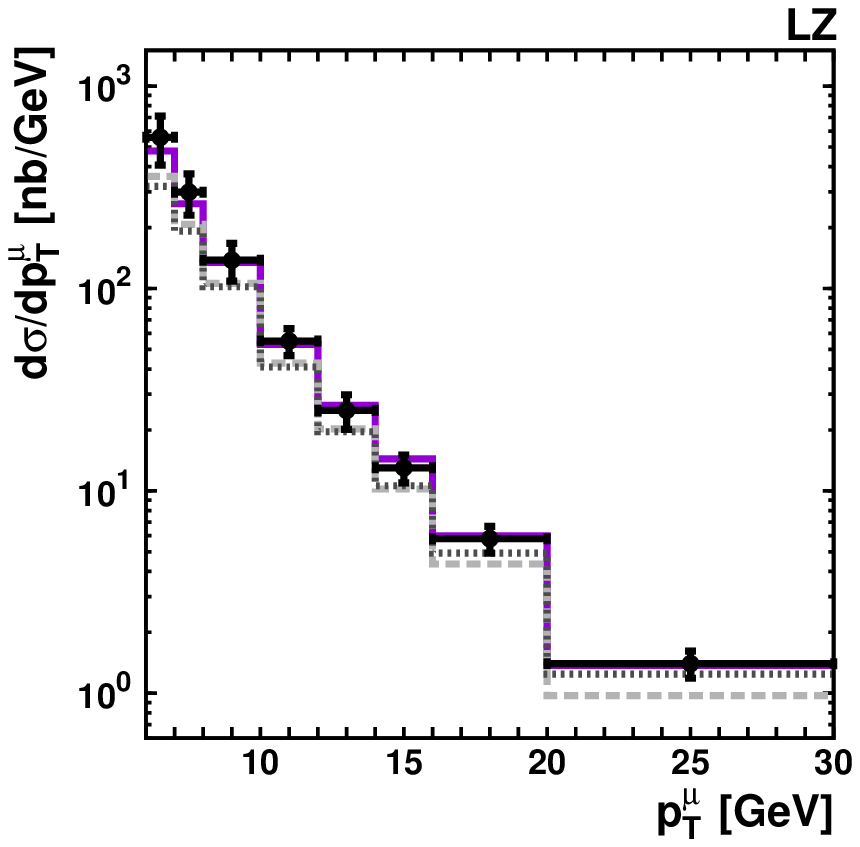, width = 8.1cm}}
\put(9.2,6.){\epsfig{figure=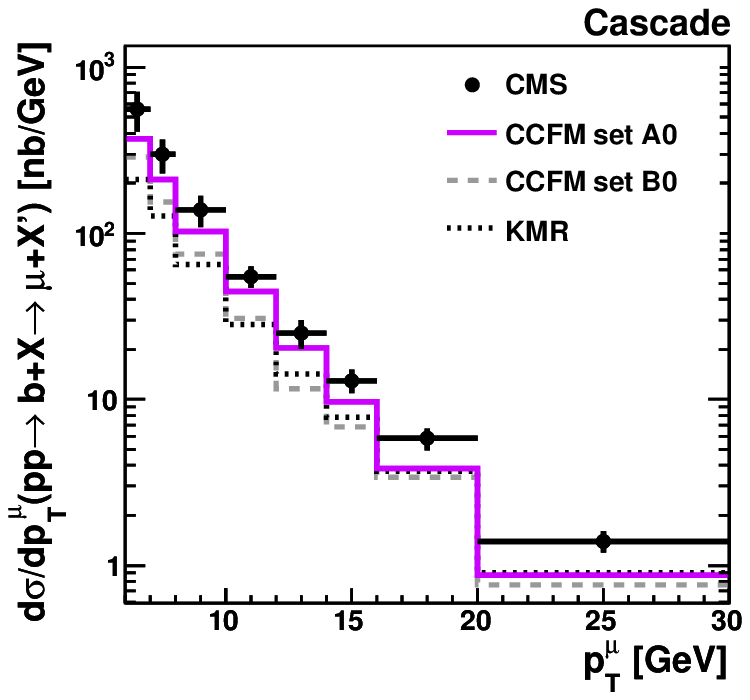, height = 6.3cm, width = 7.3cm}}
\put(0.1,0.59){\epsfig{figure=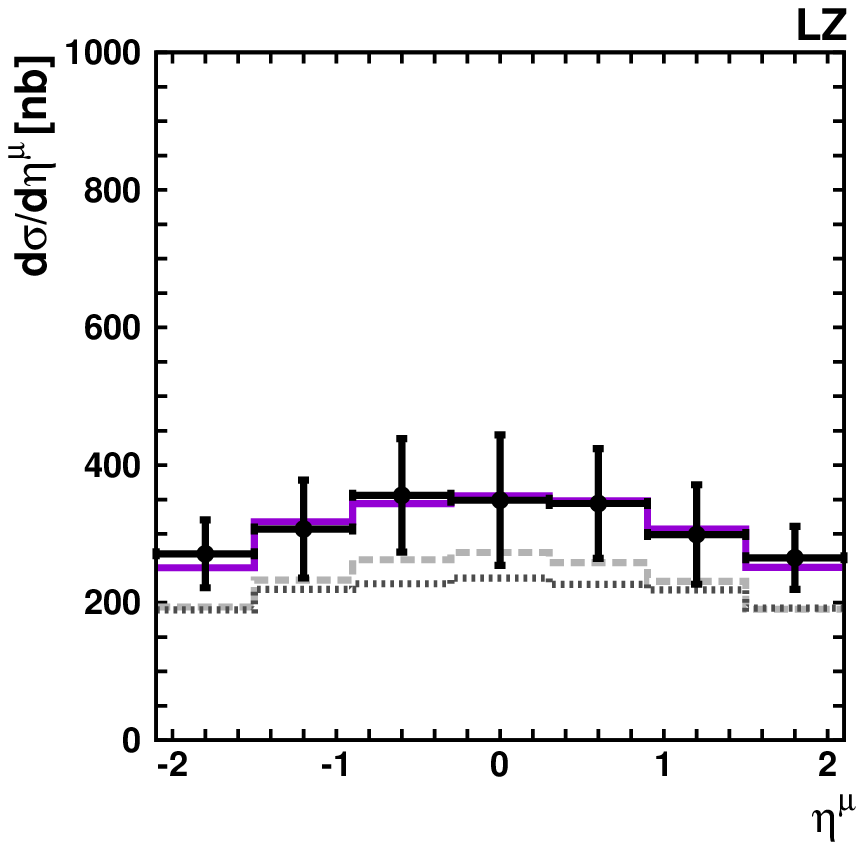, width = 8.1cm}}
\put(9.2,0.0){\epsfig{figure=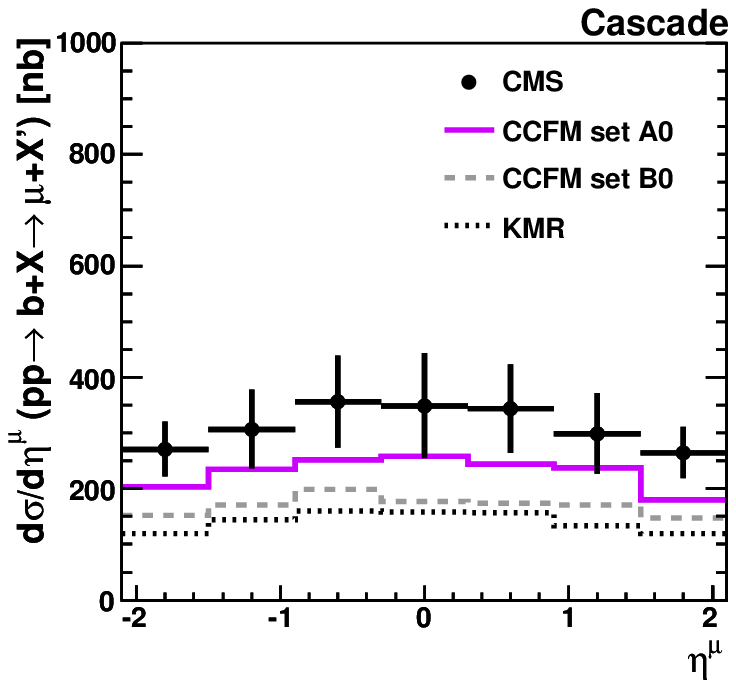, height = 6.3cm, width = 7.3cm}}
\end{picture}
\caption{The transverse momentum and pseudorapidity distributions of muons arising from the 
semileptonic decays of $b$ quarks at the LHC. The kinematical cuts applied are described in the text. 
The left histograms show 
the LZ numerical results while the right plots depict the \textsc{Cascade} predictions.
Notation of all histograms is the same as in Fig.~1. 
The experimental data are from CMS\protect\cite{4}.}
\label{fig3}
\end{figure}

\begin{figure}
\centering
\begin{picture}(16.5,7.)(0.,0.)
\put(0.1,0.4){\epsfig{figure=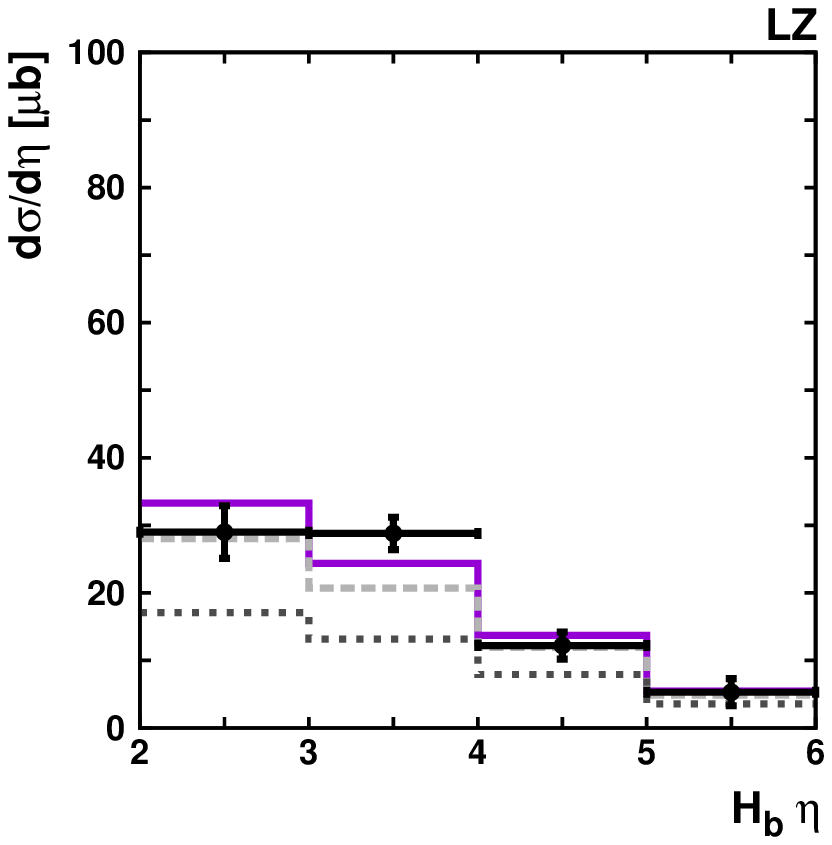, width = 8.1cm}}
\put(9.2,0.0){\epsfig{figure=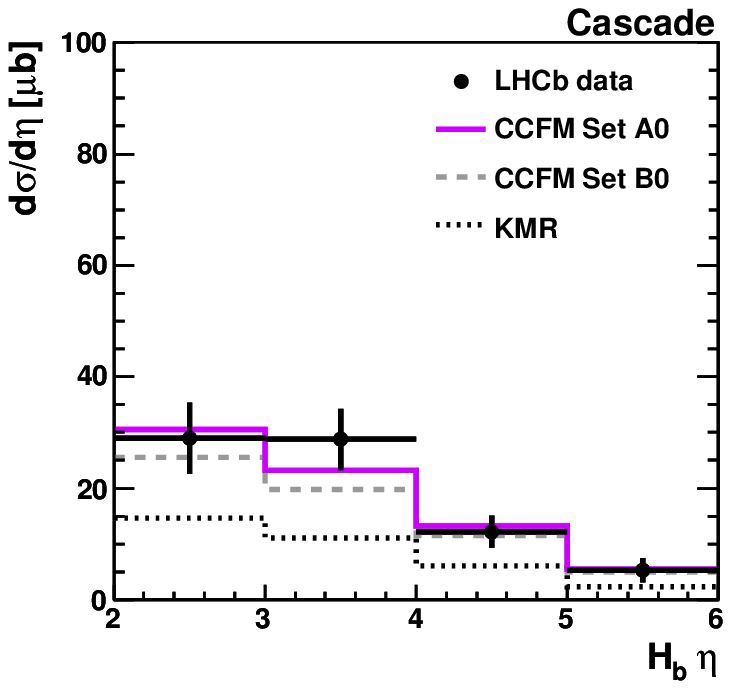, height = 6.3cm, width = 7.3cm}}
\end{picture}
\caption{The pseudorapidity distributions of $b$-flavored hadrons at the LHC. 
The kinematical cuts applied are described in the text. The left histogram shows 
the LZ numerical results while the right plot depicts the \textsc{Cascade} predictions.
Notation of all histograms is the same as in Fig.~1. 
The experimental data are from LHCb\protect\cite{1}.}
\label{fig4}
\end{figure}

\begin{figure}
\begin{center}
\epsfig{figure=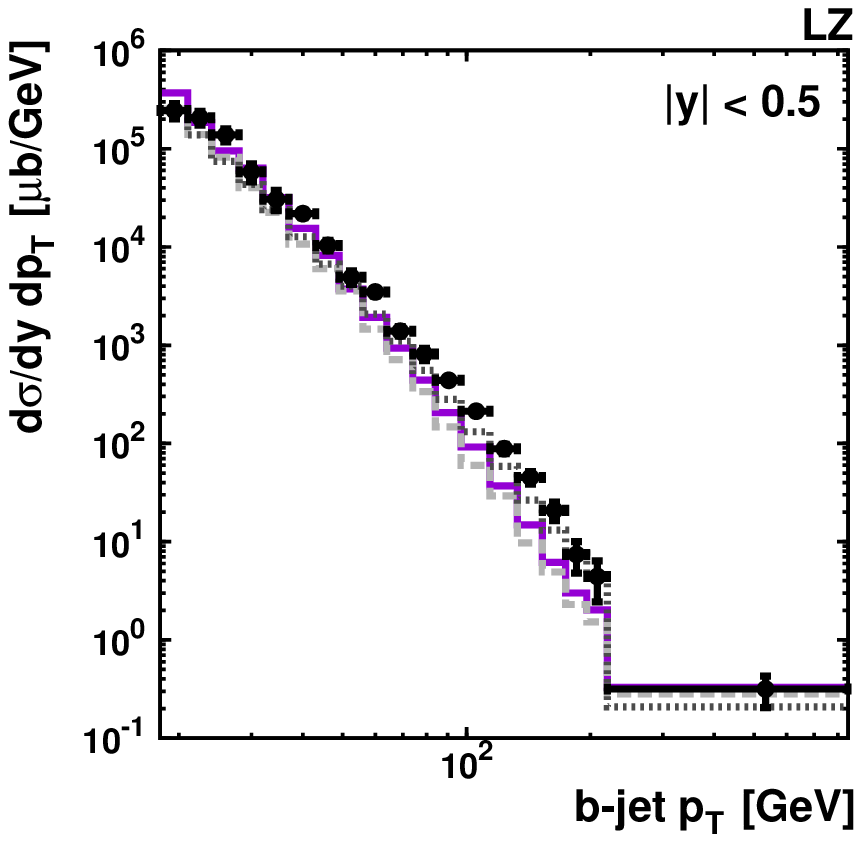, width = 8.1cm}
\epsfig{figure=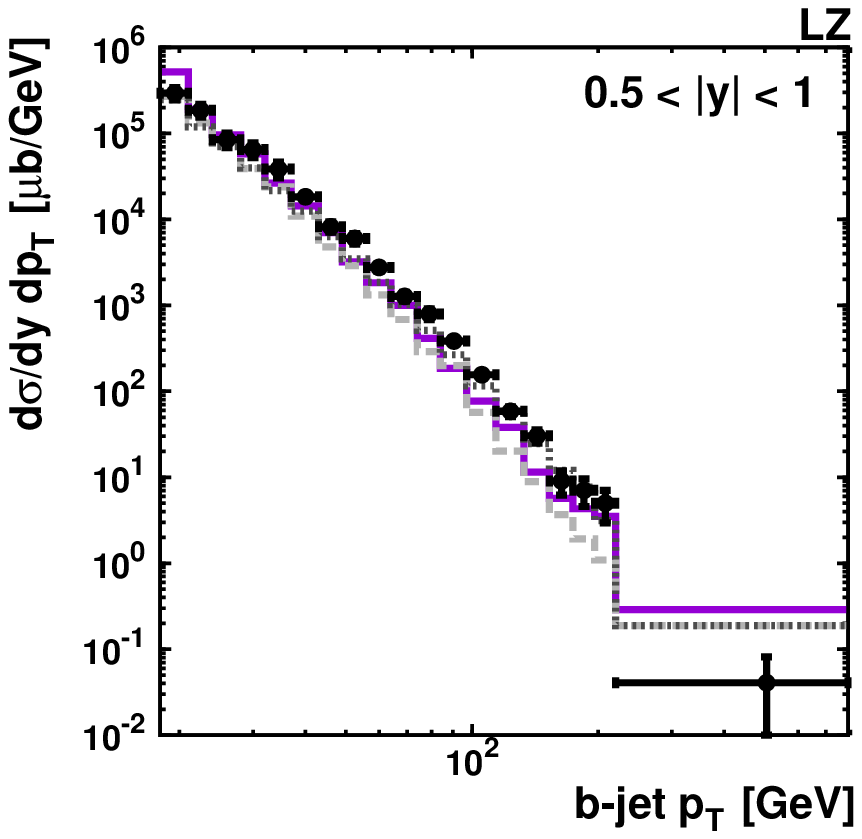, width = 8.1cm}
\epsfig{figure=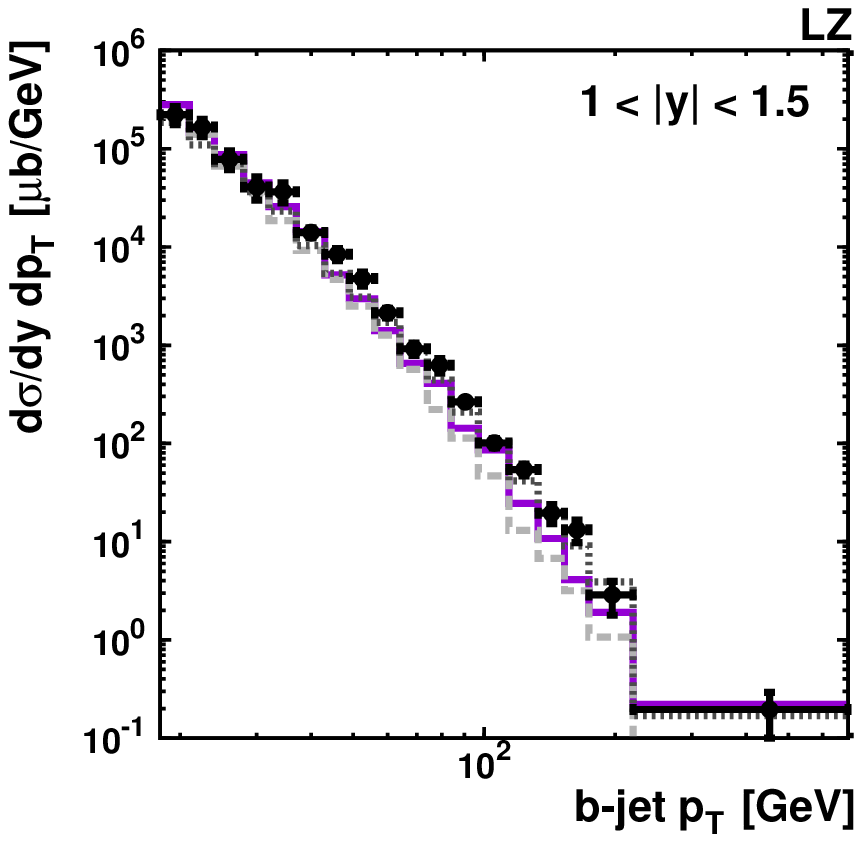, width = 8.1cm}
\epsfig{figure=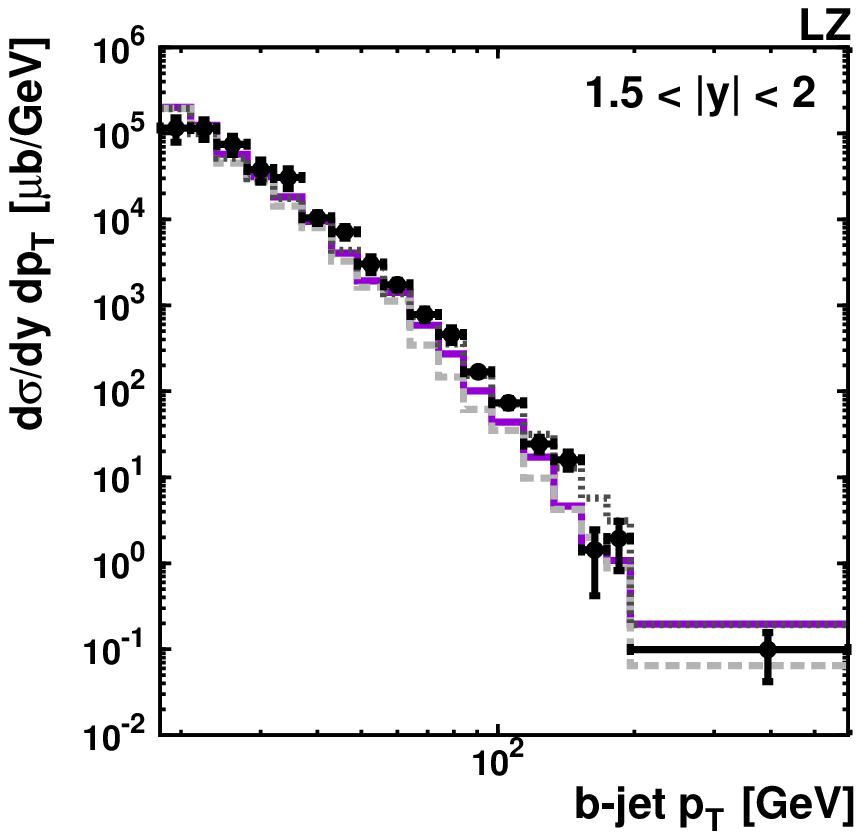, width = 8.1cm}
\caption{The double differential cross sections $d\sigma/dy\,dp_T$ of inclusive
$b$-jet production as a function of $p_T$ in different $y$ regions (LZ predictions). 
Notation of all histograms is the same as in Fig.~1.
The experimental data are from CMS\cite{6}.}
\end{center}
\label{fig5}
\end{figure}

\begin{figure}
\centering
\begin{picture}(16.5,15.)(0.,0.)
\put(1.05,6.){\epsfig{figure=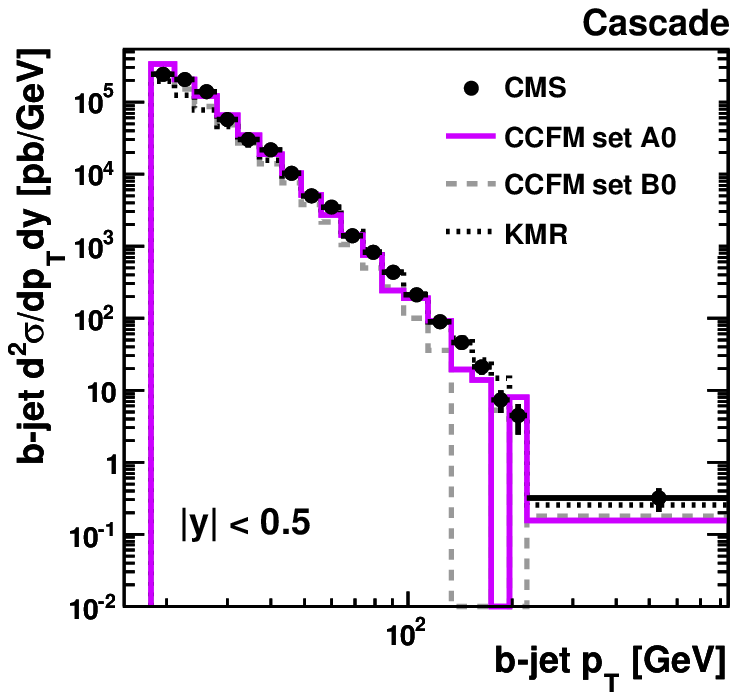, height = 6.3cm, width = 7.3cm}}
\put(9.2,6.){\epsfig{figure=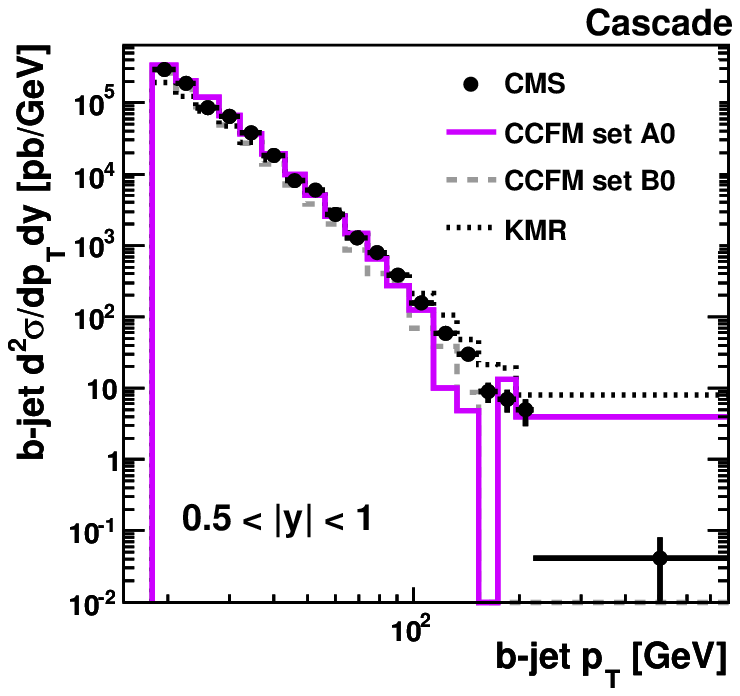, height = 6.3cm, width = 7.3cm}}
\put(1.05,0.0){\epsfig{figure=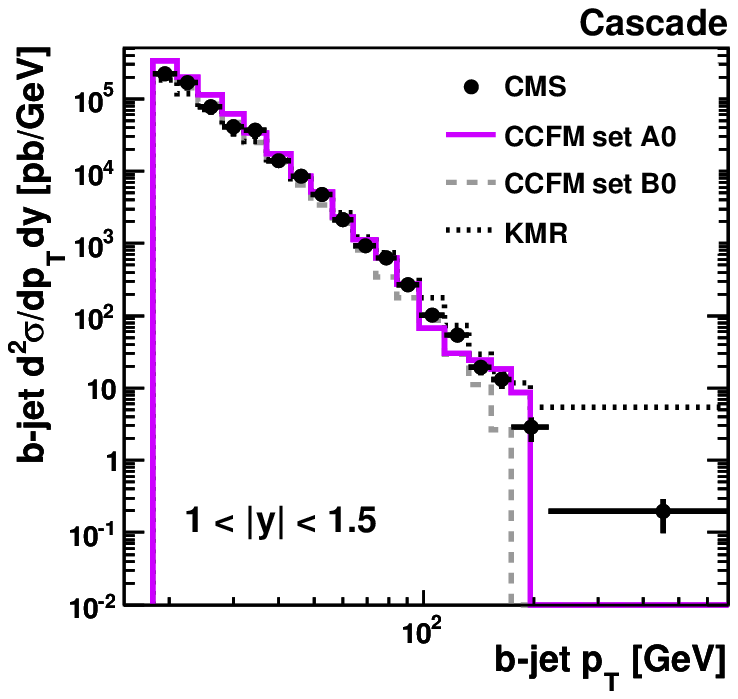, height = 6.3cm, width = 7.3cm}}
\put(9.2,0.0){\epsfig{figure=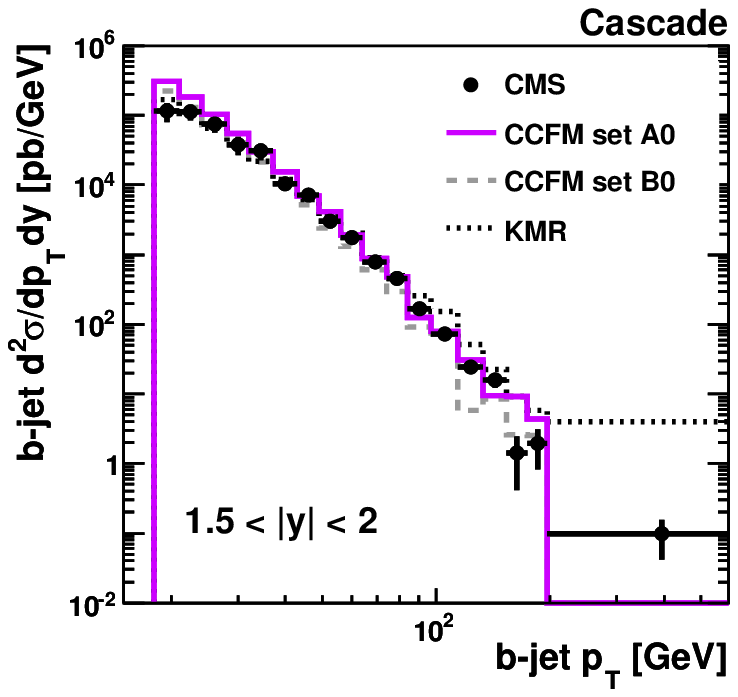, height = 6.3cm, width = 7.3cm}}
\end{picture}
\caption{The double differential cross sections $d\sigma/dy\,dp_T$ of inclusive
$b$-jet production as a function of $p_T$ in different $y$ regions
(\textsc{Cascade} predictions). 
Notation of all histograms is the same as in Fig.~1.
The experimental data are from CMS\cite{6}.}
\label{fig6}
\end{figure}

\begin{figure}
\centering
\begin{picture}(16.5,15.)(0.,0.)
\put(1.05,6.){\epsfig{figure=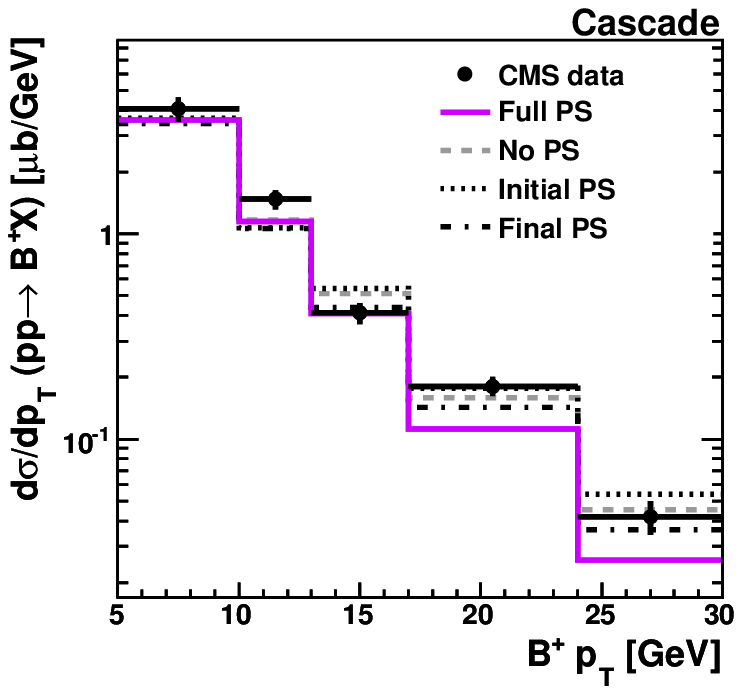, height = 6.3cm, width = 7.3cm}}
\put(9.2,6.){\epsfig{figure=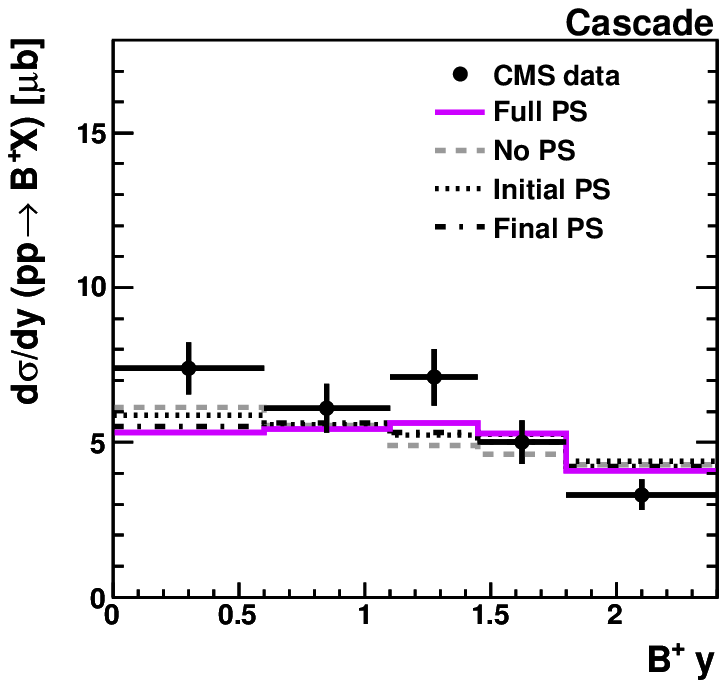, height = 6.3cm, width = 7.3cm}}
\put(1.05,0.0){\epsfig{figure=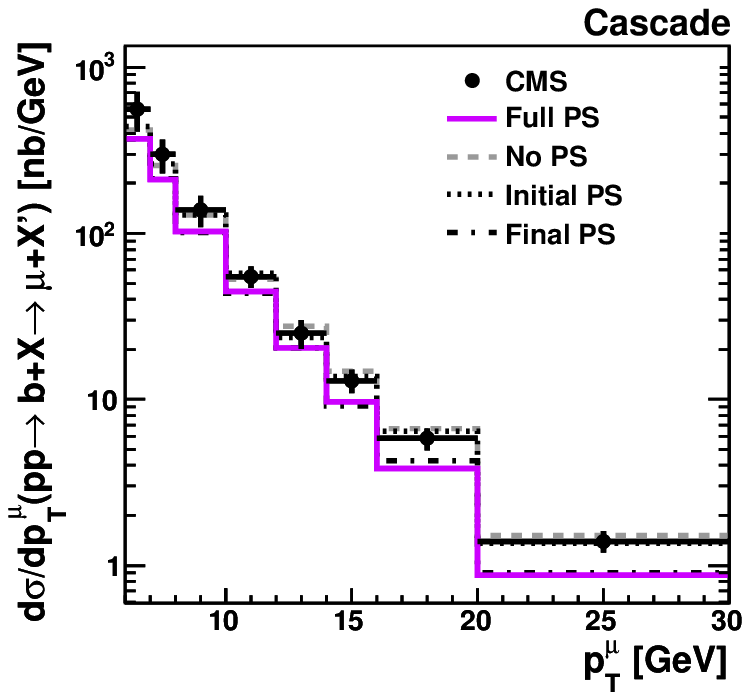, height = 6.3cm, width = 7.3cm}}
\put(9.2,0.0){\epsfig{figure=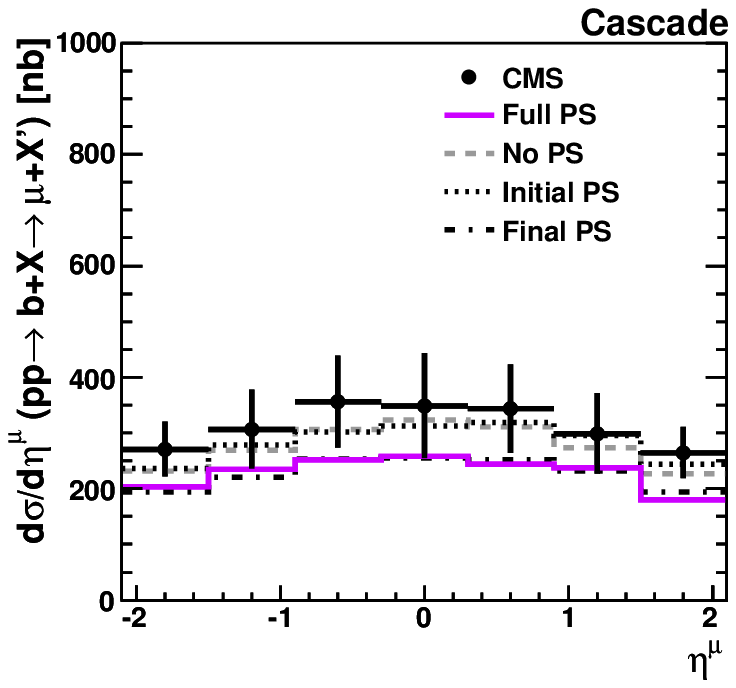, height = 6.3cm, width = 7.3cm}}
\end{picture}
\caption{Parton shower effects in the pseudorapidity and transverse momentum 
distributions of the $b$-quark decay muons. The four lines represent full 
parton shower (solid line), no parton shower (dashed line), initial state
parton shower (dotted line) and final state parton shower (dashed dotted line).
The experimental data are from CMS\cite{2,4}.}
\label{fig7}
\end{figure}

\begin{figure}
\begin{center}
\epsfig{figure=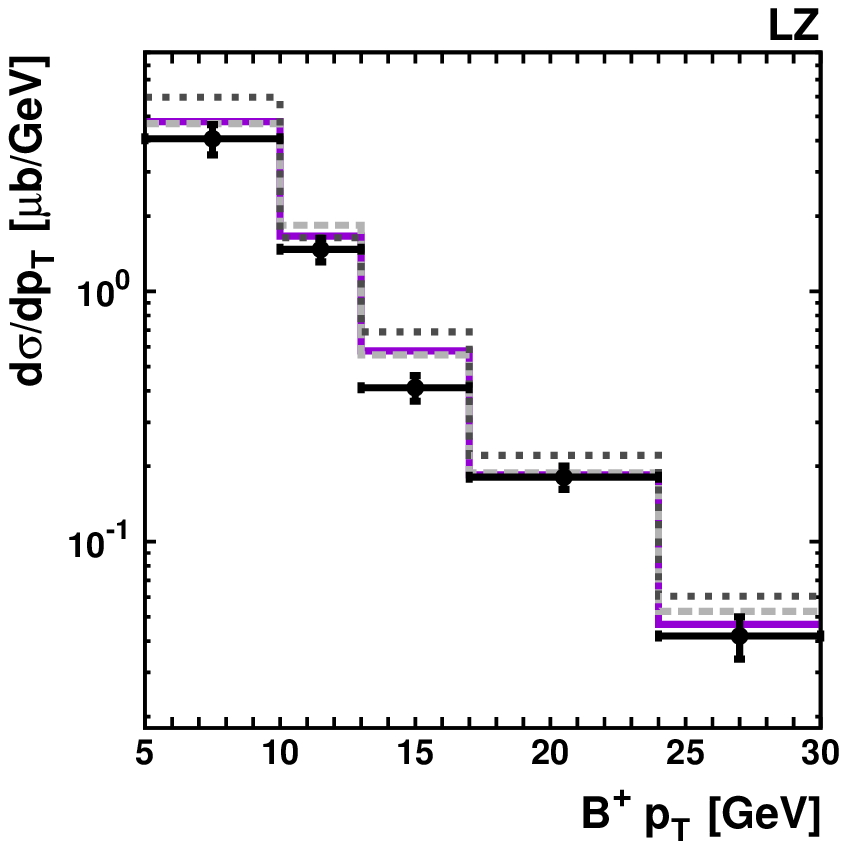, width = 8.1cm}
\epsfig{figure=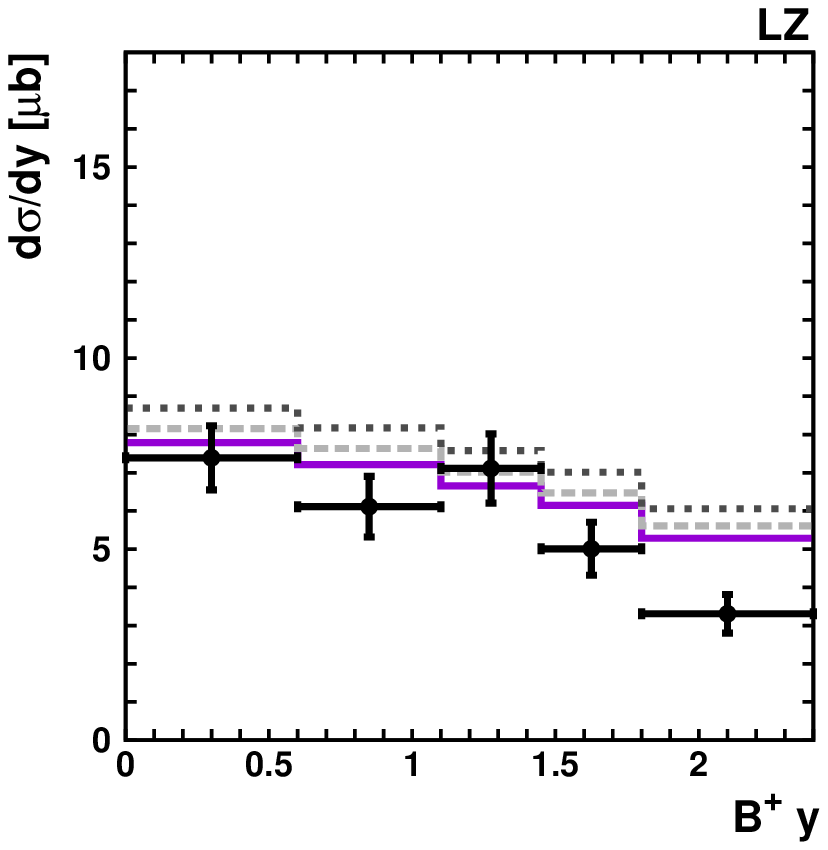, width = 8.1cm}
\epsfig{figure=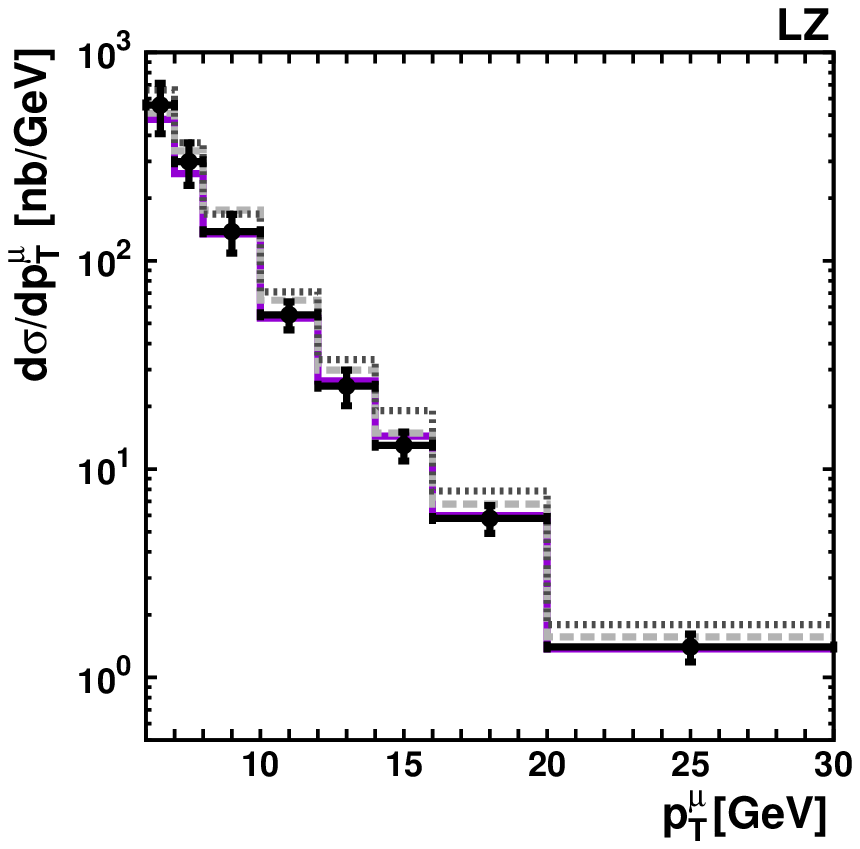, width = 8.1cm}
\epsfig{figure=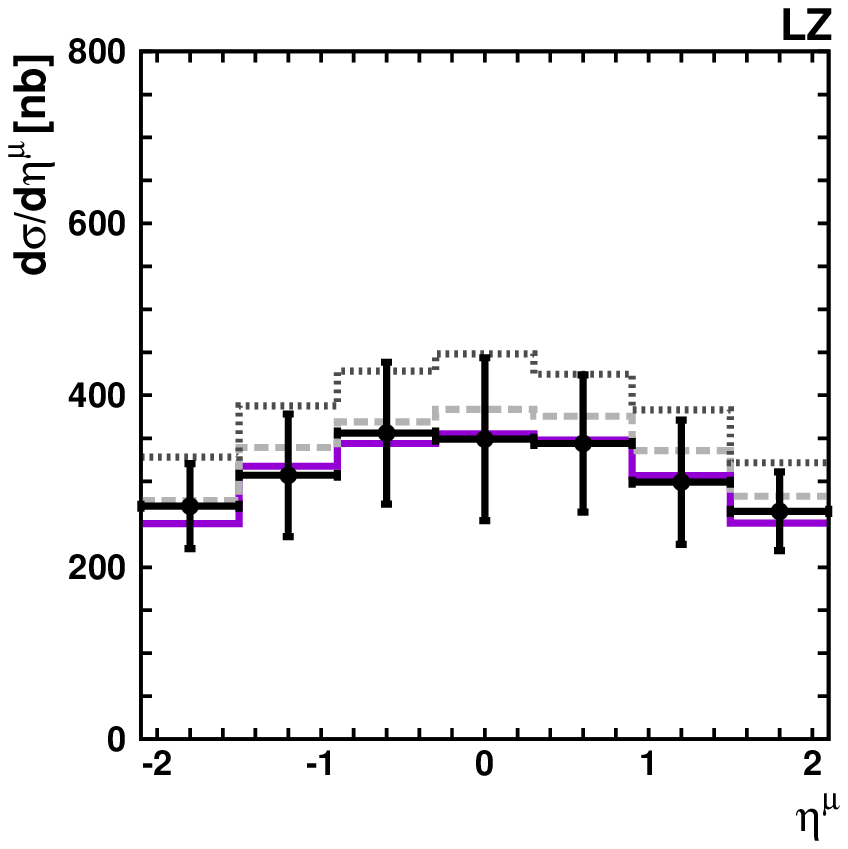, width = 8.1cm}
\end{center}
\caption{The dependence of our predictions on the fragmentation scheme (LZ calculations).
The solid, dashed and dotted histograms correspond to the results
obtained using the Peterson fragmentation function with $\epsilon_b = 0.006$, 
$\epsilon_b = 0.003$ and the non-perturbative fragmentation functions 
from\cite{23, 24, 25}, respectively. The CCFM-evolved (A0) gluon density is applied.
The experimental data are from CMS\cite{2,4}.}
\label{fig8}
\end{figure}

\begin{figure}
\begin{picture}(16.5,20.)(0.,0.)
\put(0,13.55){\epsfig{figure=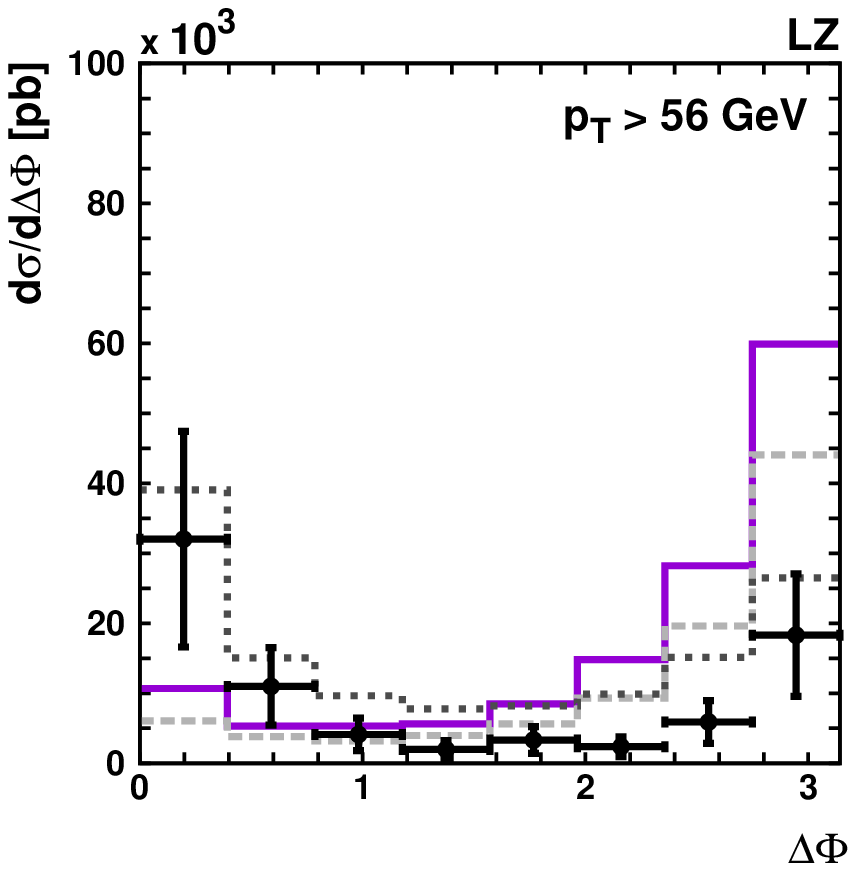, width = 8.1cm}}
\put(9.25,13){\epsfig{figure=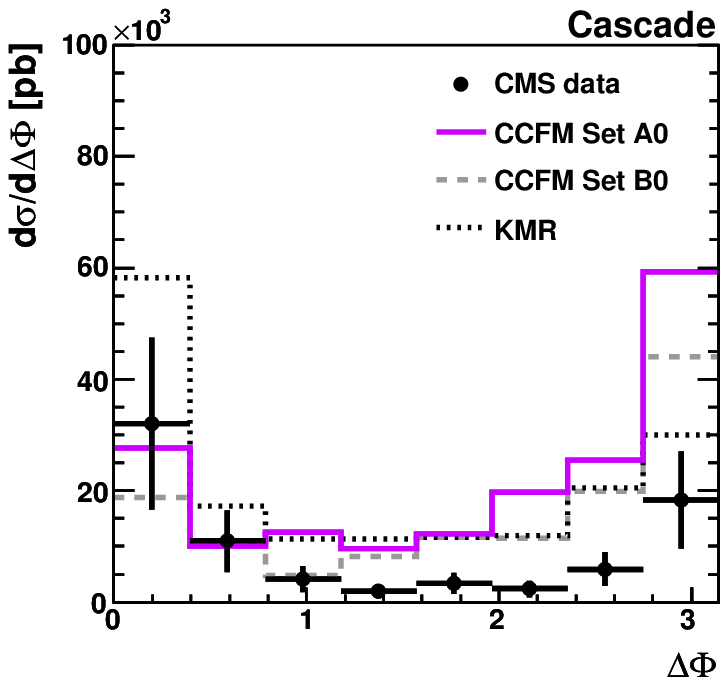, height = 6.3cm, width = 7.3cm}}
\put(0,7.2){\epsfig{figure=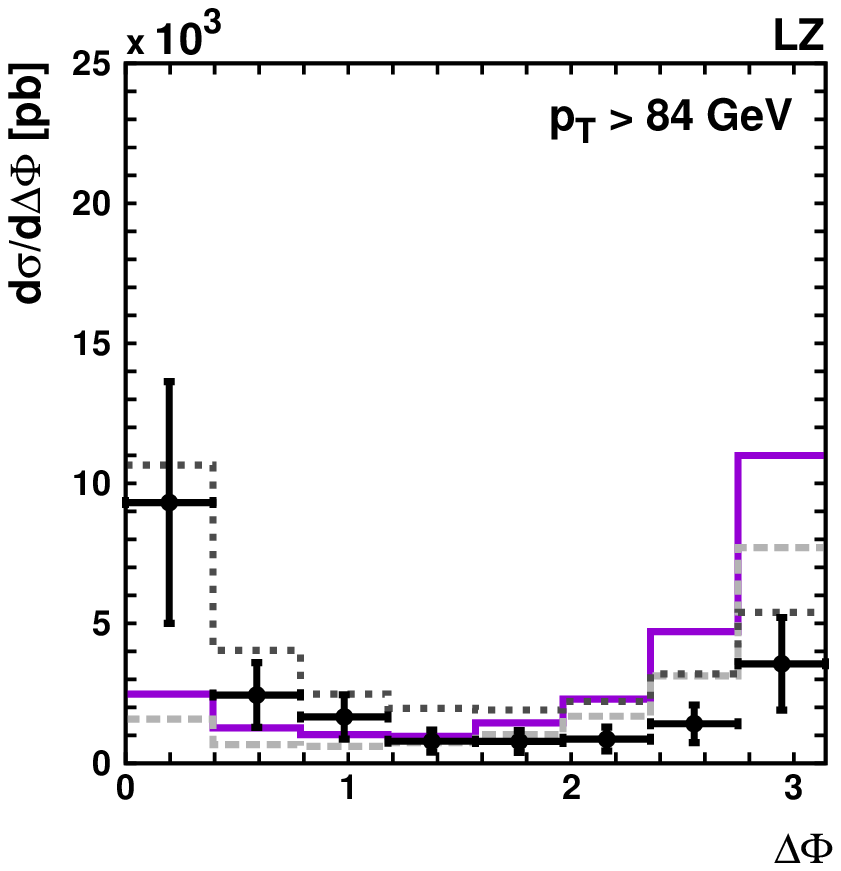, width = 8.1cm}}
\put(9.25,6.62){\epsfig{figure=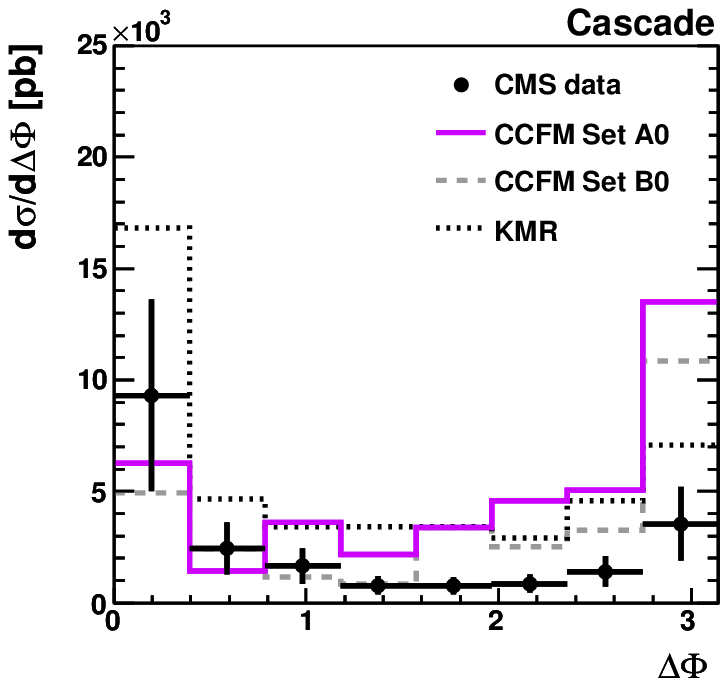, height = 6.3cm, width = 7.3cm}}
\put(0,0.55){\epsfig{figure=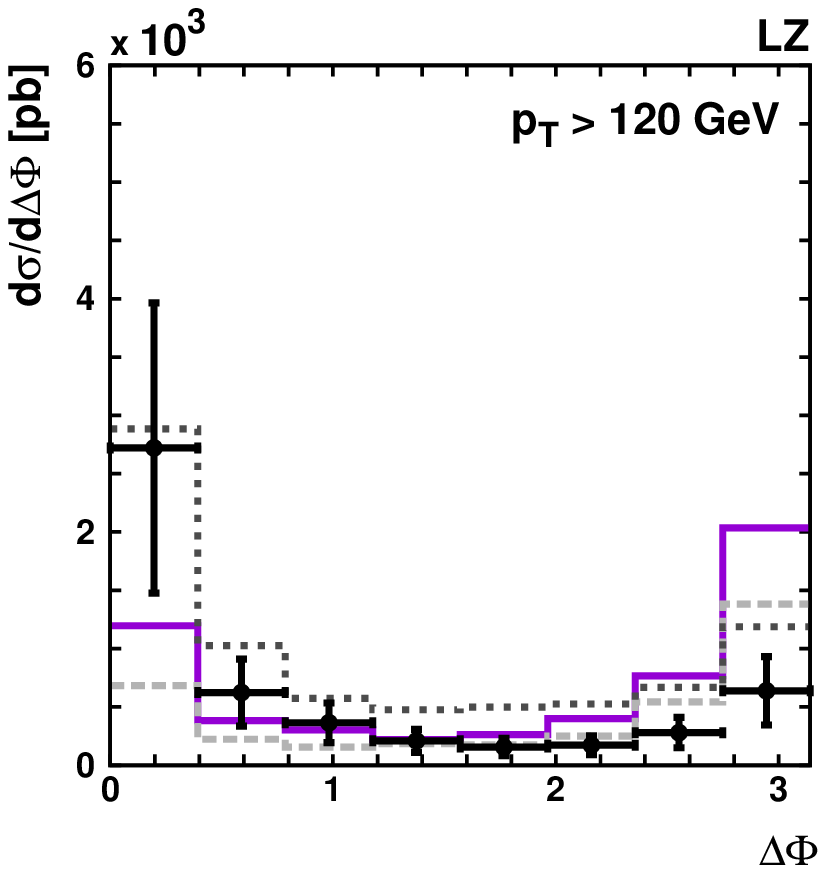, width = 8.1cm}}
\put(9.25,0){\epsfig{figure=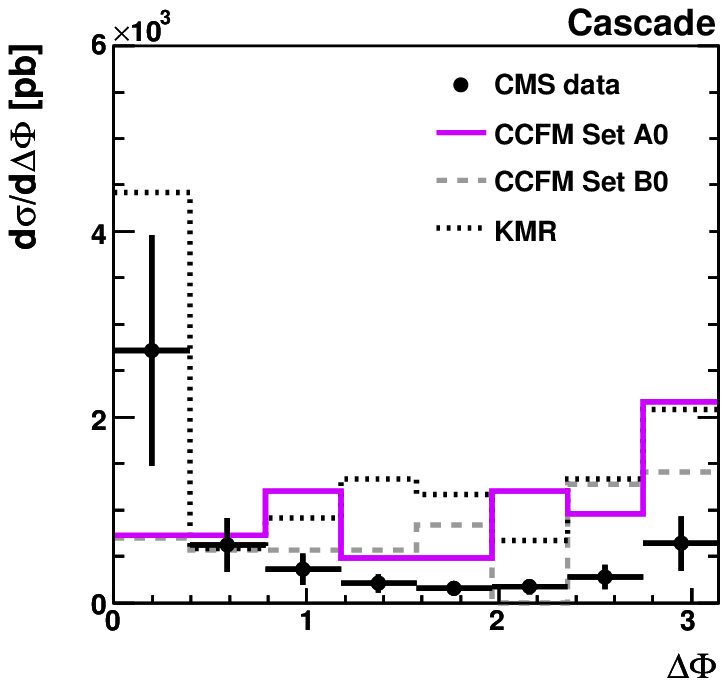, height = 6.3cm, width = 7.3cm}}
\end{picture}
\caption{The distributions in azimuthal angle difference
between the produced $b$-flavored hadrons at the LHC.
The first column shows the LZ numerical 
results while the second one depicts the \textsc{Cascade} predictions.
The kinematical cuts applied are described in the text.
Notation of all histograms is the same as in Fig.~1.
The experimental data are from CMS\cite{5}.}
\label{fig9}
\end{figure}

\begin{figure}
\begin{picture}(16.5,20.)(0.,0.)
\put(0,13.55){\epsfig{figure=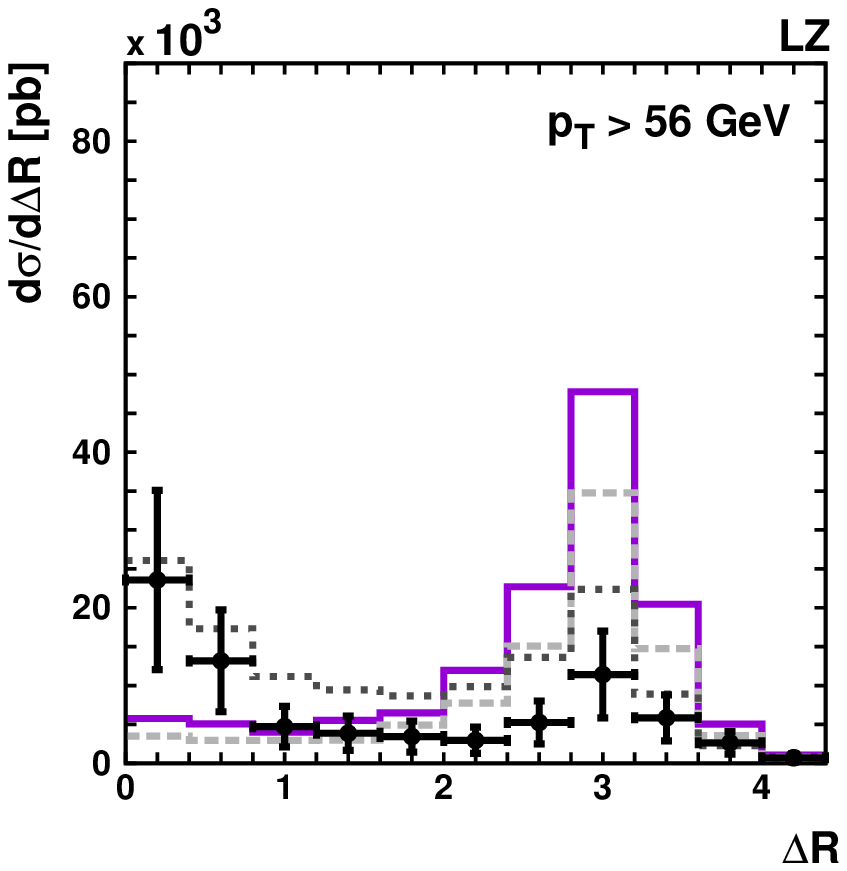, width = 8.1cm}}
\put(9.25,13){\epsfig{figure=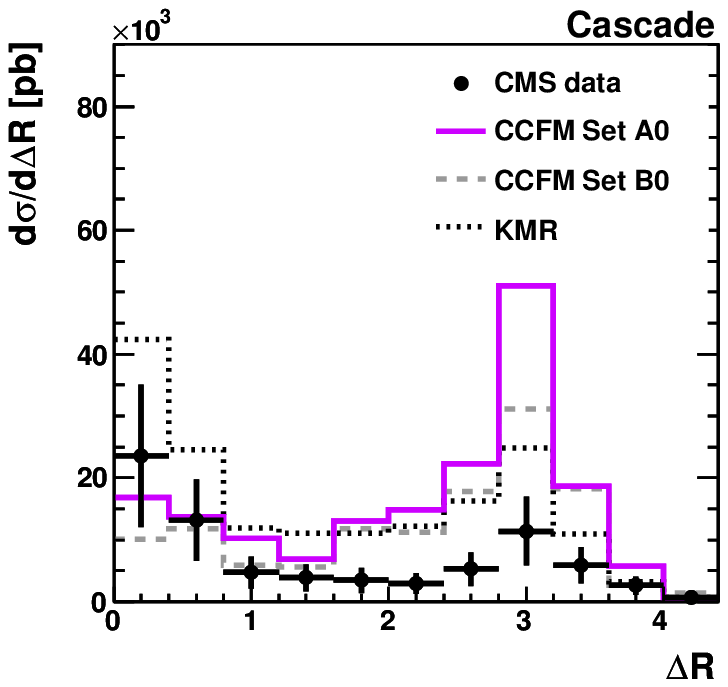, height = 6.3cm, width = 7.3cm}}
\put(0,7.2){\epsfig{figure=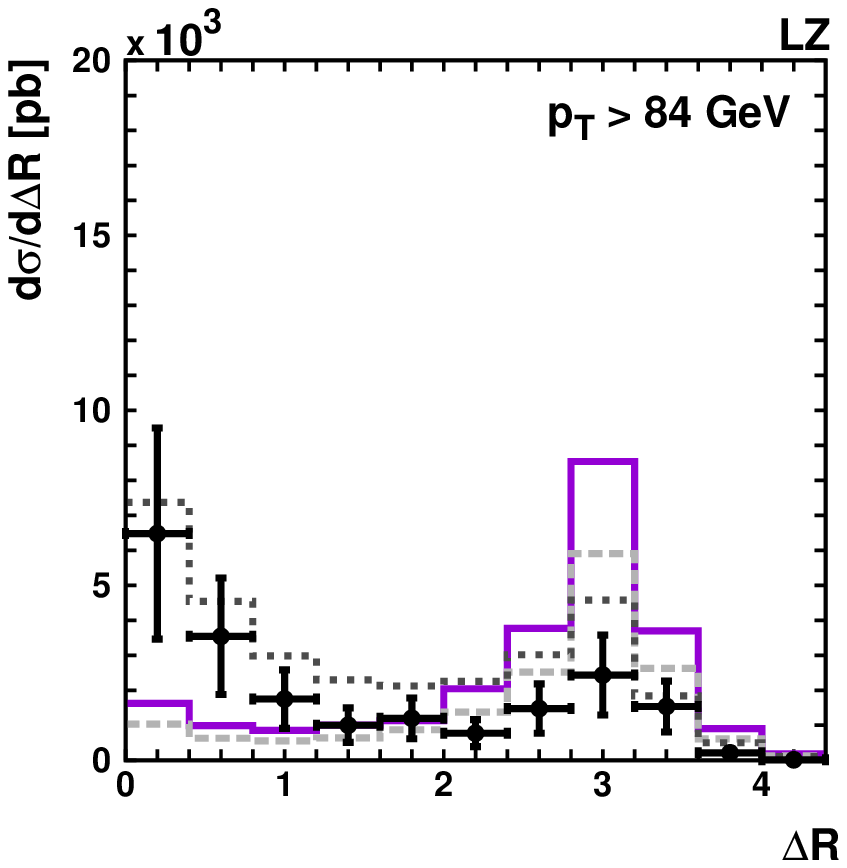, width = 8.1cm}}
\put(9.25,6.62){\epsfig{figure=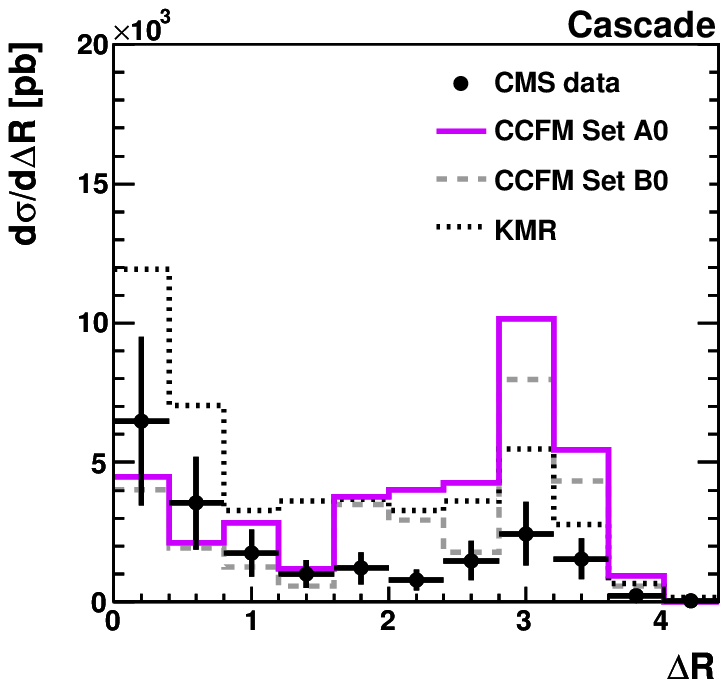, height = 6.3cm, width = 7.3cm}}
\put(0,0.55){\epsfig{figure=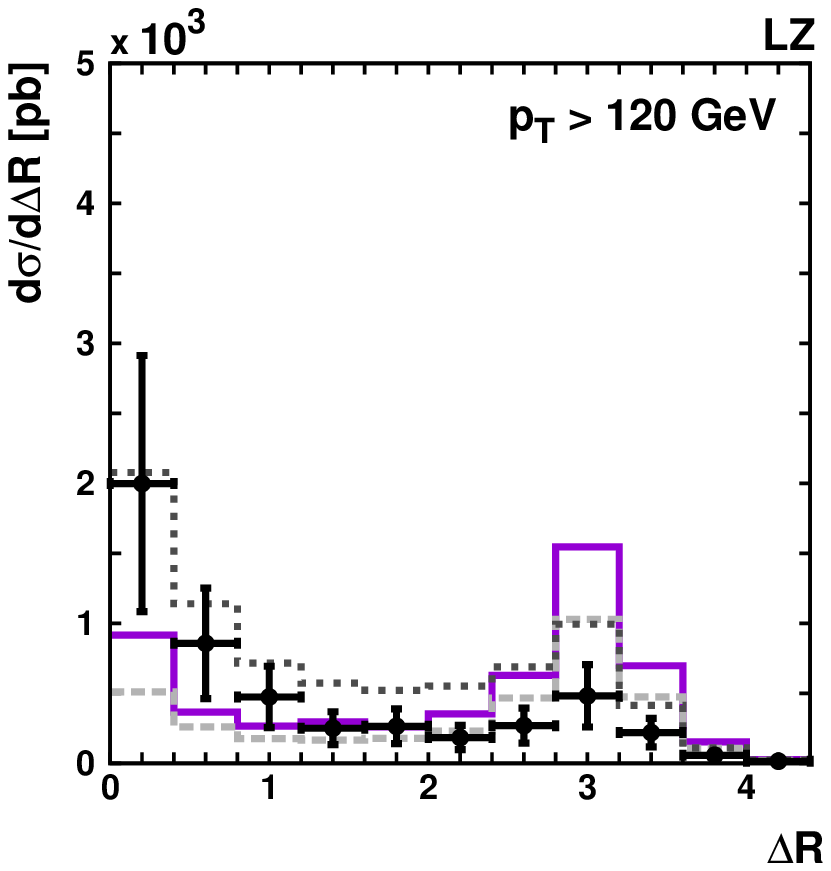, width = 8.1cm}}
\put(9.25,0){\epsfig{figure=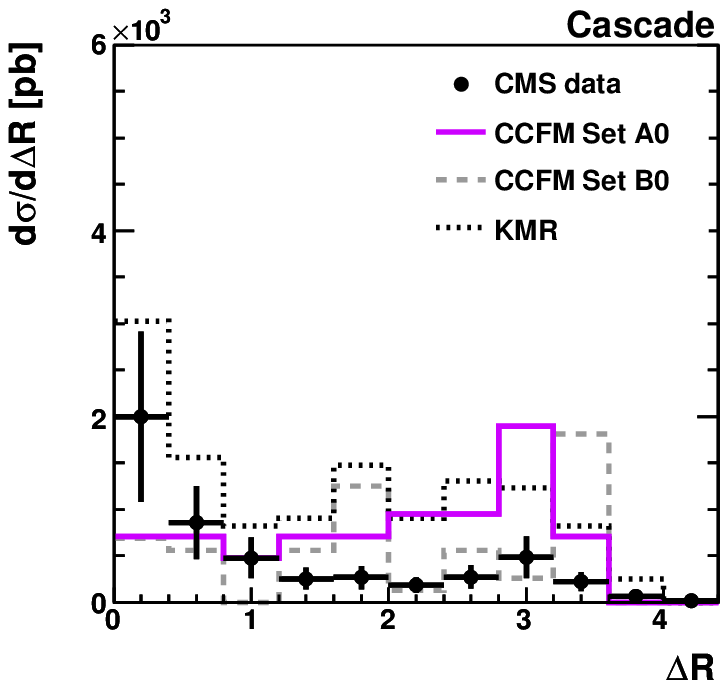, height = 6.3cm, width = 7.3cm}}
\end{picture}
\caption{The distributions in $\Delta R$ in the
$b$-flavored hadron production at the LHC.
The first column shows the LZ numerical 
results while the second one depicts the \textsc{Cascade} predictions.
The kinematical cuts applied are described in the text.
Notation of all histograms is the same as in Fig.~1.
The experimental data are from CMS\cite{5}.}
\label{fig10}
\end{figure}

\begin{figure}
\begin{center}
\epsfig{figure=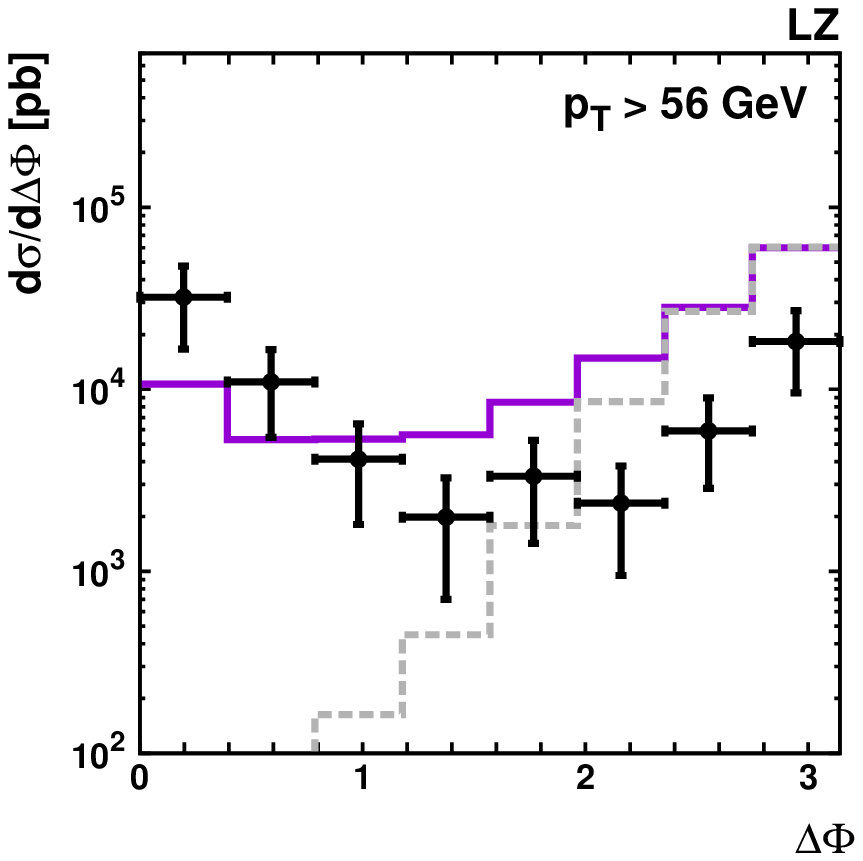, width = 8.1cm}
\epsfig{figure=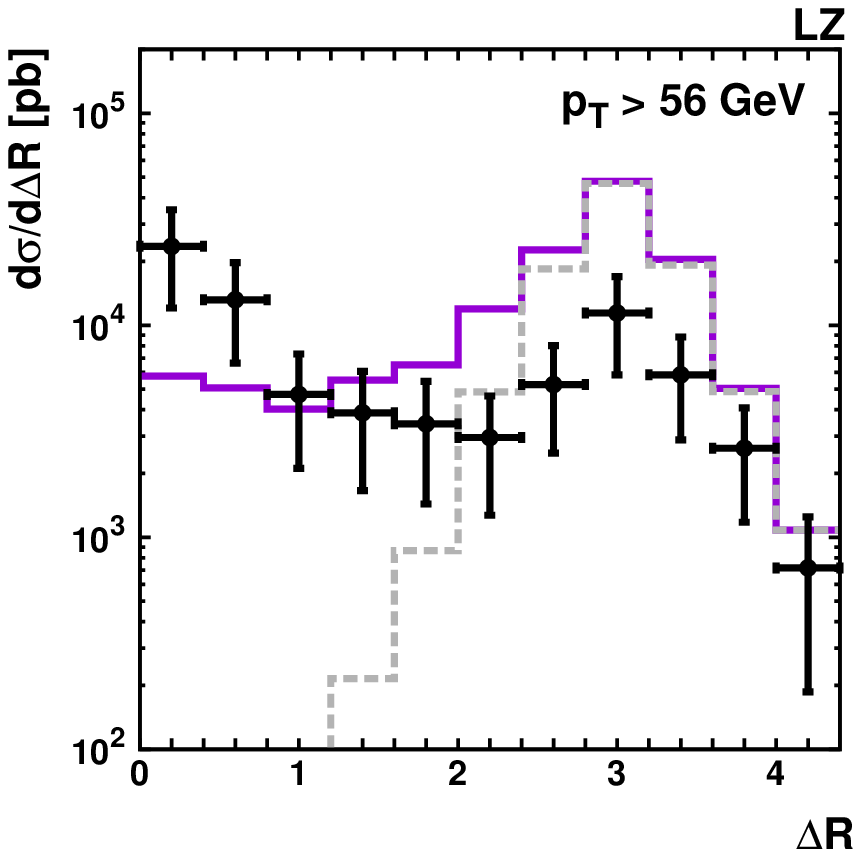, width = 8.1cm}
\epsfig{figure=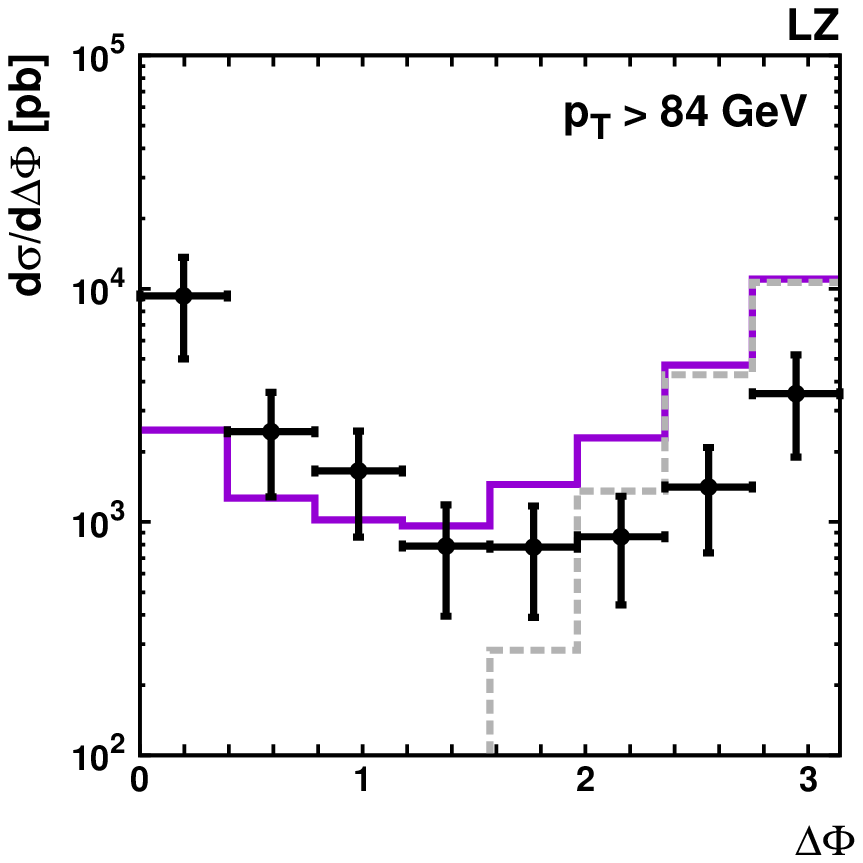, width = 8.1cm}
\epsfig{figure=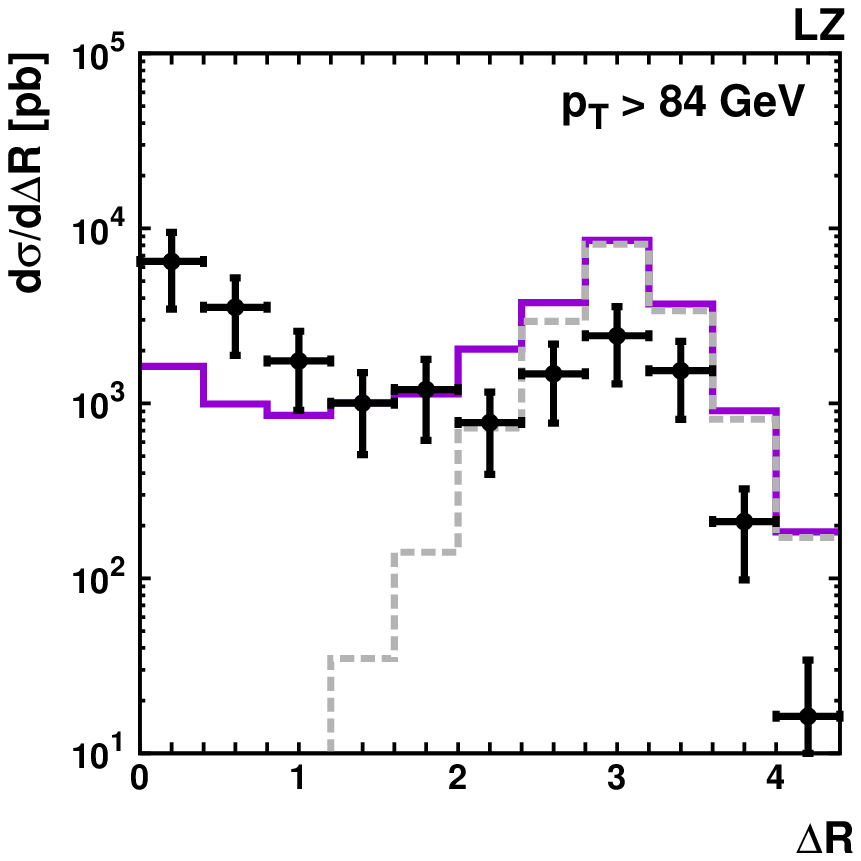, width = 8.1cm}
\epsfig{figure=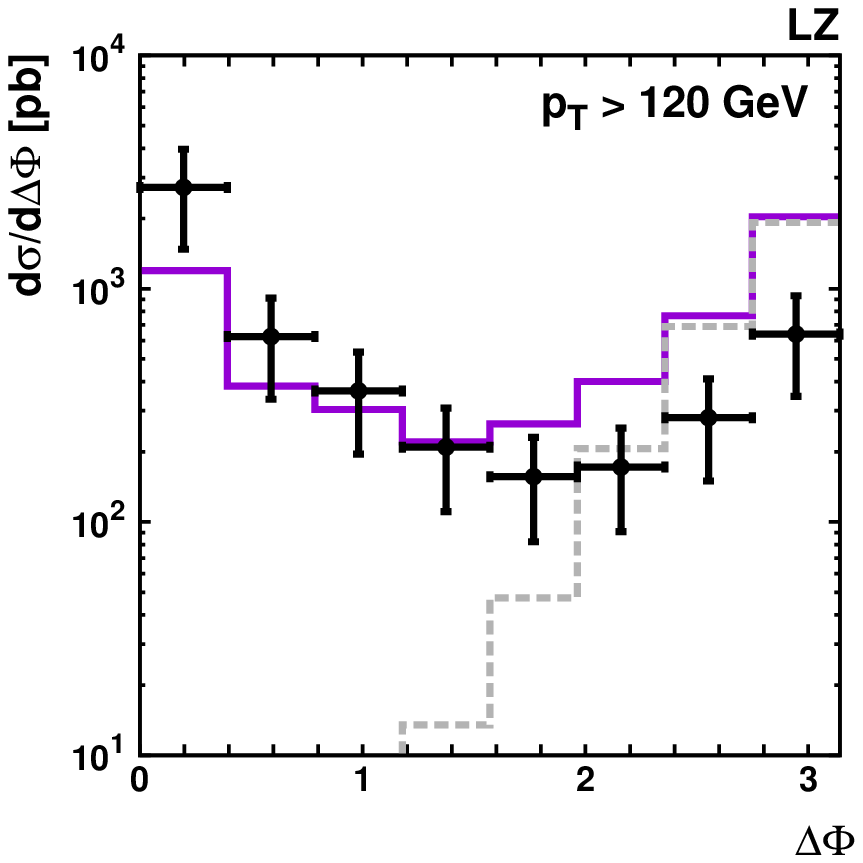, width = 8.1cm}
\epsfig{figure=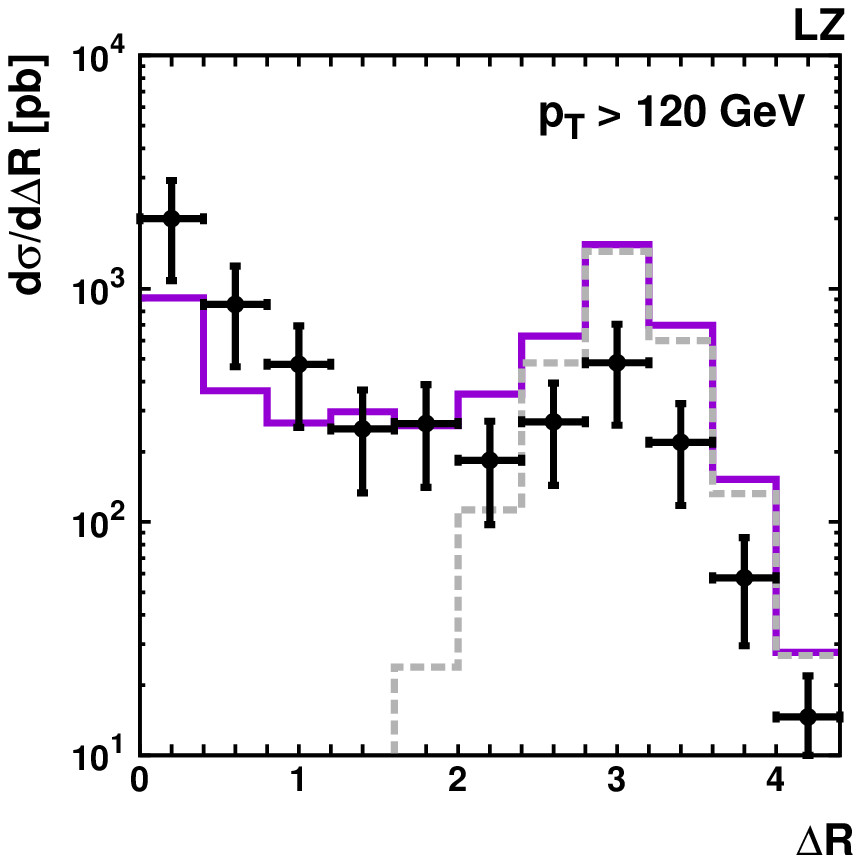, width = 8.1cm}
\caption{Importance of non-zero transverse momentum of incoming gluons in open $b$ quark production
at the LHC. The solid histograms correspond to the results 
obtained according to the master formula~(1). The dotted histograms are obtained by using 
the same formula but now we switch off the virtualities of both incoming gluons in 
partonic amplitude and apply an additional requirement ${\mathbf k}_{1,2 \,T}^2 < \mu_R^2$.
We have used here the CCFM A0 gluon for illustration.
The experimental data are from CMS~\cite{5}.}
\end{center}
\label{fig11}
\end{figure}

\end{document}